\pdfoutput=1
\documentclass[prb,showkeys,preprintnumbers,amsmath,amssymb,nofootinbib,10pt,twocolumn,altaffilsymbol,final]{revtex4-2}
\usepackage[english]{babel}
\usepackage[T1]{fontenc}
\usepackage{graphicx}
\usepackage{hyperref}
\usepackage{xspace}
\newcommand*{\ggll}{\ensuremath{\gamma\gamma\rightarrow\ell^{+}\ell^{-}}\xspace}

\newcommand{\kt}{\ensuremath{k_{\rm T}}\xspace}
\newcommand{\pt}{\ensuremath{p_{\perp}}\xspace}
\newcommand{\gev}{\ensuremath{\rm GeV}\xspace}

\newcommand{\qsel}{\ensuremath{Q_{\rm el}^2}\xspace}
\newcommand{\qsinel}{\ensuremath{Q_{\rm inel}^2}\xspace}
\newcommand{\veckt}{\ensuremath{{\bf k}_{\perp}}\xspace}
\newcommand{\xbj}{\ensuremath{x_{\rm Bj}}\xspace}

\newcounter{bla}

\begin{document}

\title{CepGen -- A generic central exclusive processes event generator for hadron-hadron collisions}
\author{Laurent \surname{Forthomme}}
\email{laurent.forthomme@cern.ch}
\altaffiliation[Now at ]{European Organization for Nuclear Research (CERN), Geneva, Switzerland.}
\affiliation{Helsinki Institute of Physics, University of Helsinki, Finland}

\begin{abstract}
  We present an event generator for the simulation of central exclusive processes in hadron-hadron reactions.
  Among others, it implements the two-photon production of lepton pairs previously introduced in LPAIR.
  As a proof of principle, we show that the two approaches are numerically consistent.
  The \kt-factorized description of this process is also handled, along with the two-photon production of a quark, or a $W^\pm$ gauge boson pair.
  This toolbox may be used as a common framework for the definition of many other processes following this approach.
  Additionally, photoproduction and other photon induced processes are also considered, or being implemented.
\end{abstract}


\maketitle

\section{Introduction}
We present a new event simulation tool to facilitate the study of central exclusive processes (CEP) in the scope of high-energy colliding experiments.

Lately, with the experimental observations of photon-induced processes at TeV scale energies with or without the help of forward proton taggers, the need for such a common simulation framework that would allow both the phenomenology and experimental studies of this specific class of events, and easily manageable by the two communities, has been clearly motivated.

In particular, the two-photon production of lepton pairs in proton-proton reactions through a $t$-channel exchange will be covered here.
This choice may be justified by the well demonstrated theoretical interest of this process, namely the precision attached to its purely elastic predictions (a pure tree-level electrodynamic component), and large uncertainties to inelastic contributions.
The latter are closely related to the proton's electromagnetic and nuclear structure modeling, thus enabling its precision study in a hadron-hadron collider environment.

Unlike the vast majority of common CEP event generators implementing the incoming photon fluxes through the equivalent photon approximation (EPA) \cite{Budnev:1974de}, the full unintegrated photon virtuality (both the transverse and longitudinal components) information is used here.
This allows a refined treatment of processes in which low-virtualities are accounting for the biggest fraction of the total cross section, such as the two-photon process quoted above.

We will therefore concentrate on this particular photon-induced process as a proof of principle for the operation and validation of this simulation tool.
~\\

This paper is organized as follows: in Section \ref{sect:phys-case} we enumerate and describe some of the processes currently implemented in this code.
Then, in Section \ref{sect:implem} we present a review of the technicalities carried in its implementation.
In Section \ref{sect:additions} we list the extensions currently embedded in CepGen to ease the user interaction and allow an increased modularity with its core features.
Finally, our conclusions are summarized in Section \ref{sect:conclusion}.

\section{Physics cases}
\label{sect:phys-case}

Beside the two-photon ($t$-channel) production of lepton (or fermion) pairs already mentioned in this paper, multiple matrix elements are included in the current version of our event generator.
For instance, one may quote:

\begin{itemize}
\item the two-photon production of a $W^\pm$ gauge boson pair, recently studied experimentally at the LHC \cite{Chatrchyan:2013akv,Khachatryan:2016mud,Aaboud:2017oiq}, and formulated in \cite{Luszczak:2018ntp} ;
\item the diffractive photoproduction of low- and intermediate mass vector mesons, previously implemented in DIFFVM \cite{List:1998jz} ;
\item the inclusive, gluon-induced production of a fermion pair, using unintegrated parton distributions following the KMR procedure \cite{Kimber:2001sc,Watt:2003vf}.
\end{itemize}

A detailed review of the last two processes will appear in a future paper.

If we again concentrate on the two-photon production of a fermion pair, two approaches are implemented within CepGen.
For the dilepton final state, one may find a modern retranscription of the LPAIR code \cite{Vermaseren:1982cz,Baranov:1991yq} used intensively in various HERA searches, and introduced in Section \ref{sect:matrixelement-lpair}.
Along with the two-photon production of $W^\pm$ pairs, it can also be formulated through the \kt-factorization technique described in \cite{daSilveira:2014jla,Luszczak:2015aoa}.

The elastic photon emission through this \kt-factorized process may also be formulated for heavy-ion initial beams.
In CepGen, this initial state is still being validated while this paper is released.

\subsection{LPAIR matrix element}
\label{sect:matrixelement-lpair}

The photon emission from initial state interacting protons may be divided in two main sub-components.
For the single-, double-dissociative (or semi-exclusive, non-exclusive) scenarios of the two-photon production of leptons pair, the EPA is not sufficient to reproduce the high decorrelation of the two individual leptons observed in the elastic case.

In the scope of an $ep$ collider such as HERA, the LPAIR event generator, with its full two-to-four matrix element definition and its implementation of the proton structure functions modeling, enabled the probe of this additional final state in which the proton would scatter a photon with a sufficient virtuality to excite and fragment in its final state.
The strong benefit of LPAIR was its strong numerical treatment of the phase space, ensuring a good stability of its matrix element despite the low-$Q^2$ range of the incoming state photons accounting for a large fraction of the total cross section.
This feature was particularly relevant in the elastic case.

The version implemented in CepGen handles by default the $pp$ initial state.
It can however be steered to retrieve the original $ep$ case.
Originally interfaced to the JETSET library, the diffractive/dissociated proton(s) could be studied in terms of its/their decay products in LPAIR.
Based on this approach, a refreshed implementation of this fragmentation feature has been ported to CepGen, as discussed in Section \ref{sect:ext-hadr}.

\subsection{\kt-factorized matrix elements}
\label{sect:matrixelement-pptoll}

As previously introduced in \cite{daSilveira:2014jla} for two-photon processes, the \kt factorization approach allows to treat separately the hard process definition and the modeling of incoming photon fluxes.
Unlike the collinear approximation, photons densities are left unintegrated and their transverse momentum contribution can be accounted for in the central kinematic variables description.

The differential cross section of a generic $pp\to p^{(\ast)}(\gamma\gamma\to X)p^{(\ast)}$ (here, $p^{(\ast)}$ conventionally expresses both the final state protonic system after and elastic or dissociative emission of the photon) can therefore be expressed as follows:
\begin{widetext}
\begin{equation}\label{eq:factorisation}
\begin{aligned}
  \frac{{\rm d}\sigma_{pp\to p^{(\ast)}Xp^{(\ast)}}}{{\rm d}\Omega}
  =\int\frac{{\rm d}^2{\veckt}_{,1}}{\pi{\veckt}_{,1}} F_{\rm el,inel}^{\gamma}(\xi,{\veckt^2}_{,1})
  \int\frac{{\rm d}^2{\veckt}_{,2}}{\pi{\veckt}_{,2}} F_{\rm el,inel}^{\gamma}(\xi,{\veckt^2}_{,2})
  \frac{{\rm d}\hat\sigma_{\gamma\gamma\to X}}{{\rm d}\Omega},
\end{aligned}
\end{equation}
\end{widetext}
with $F_{\rm el,inel}^{\gamma}$ the unintegrated incoming photon fluxes, and ${\rm d}\hat\sigma$ the differential cross section for the transverse momentum dependent two-photon process.
The earlier may be expressed as a function of the fractional proton momentum $\xi$ carried by the scattered photon, and its squared transverse momentum norm $\veckt^2$.

For instance, using the Budnev prescription quoted in \eqref{eq:factorisation}, one may define the following modeling for the low-$Q^2$ range elastic scattering where the proton remains on-shell after photon emission:
\begin{widetext}
\begin{equation}\label{eq:budnev_el}
\begin{aligned}
  F_{\rm el}^{\gamma}(\xi,\veckt^2) = \frac{\alpha}{\pi}\left[(1-\xi)\left(\frac{\veckt^2}{\veckt^2+\xi^2 m_p^2}\right)^2 F_E(\qsel)
  +\frac{\xi^2}{4}\left(\frac{\veckt^2}{\veckt^2+\xi^2 m_p^2}\right) F_M(\qsel)\right],
\end{aligned}
\end{equation}
\end{widetext}
where \qsel, the photon virtuality for elastic emission is obtained from the fractional proton momentum loss and the transverse component photon virtuality:
\begin{displaymath}
\qsel(\xi,\veckt^2)=\frac{\veckt^2+\xi^2 m_p^2}{1-\xi}.
\end{displaymath}
In the formalism of \eqref{eq:budnev_el}, we use:
\begin{eqnarray*}
F_E(Q^2) = \frac{4m_p^2 G_E^2+Q^2 G_M^2}{4m_p^2+Q^2},\hspace{1em}
F_M(Q^2) = G_M^2,
\end{eqnarray*}
as linear combinations of $G_{E,M}(Q^2)$, the electric and magnetic form factors of the proton.

As for the unintegrated inelastic scattering flux, where the initial proton loses a sufficiently high virtuality to be excited into a dissociative system of mass $M_X$, it can be expressed as:
\begin{widetext}
\begin{equation}\label{eq:budnev_inel}
\begin{aligned}
  F_{\rm inel}^{\gamma}(\xi,\veckt^2)
  =\frac{\alpha}{\pi}\Bigg[(1-\xi)\left(\frac{\veckt^2}{\veckt^2+\xi(M_X^2-m_p^2)+\xi^2 m_p^2}\right)^2\frac{F_2(\xbj,\qsinel)}{\qsinel+M_X^2-m_p^2}+{}\\
  {}+\frac{\xi^2}{4}\frac{1}{\xbj^2} \left(\frac{\veckt^2}{\veckt^2+\xi(M_X^2-m_p^2)+\xi^2 m_p^2}\right) \frac{2\xbj F_1(\xbj,\qsinel)}{\qsinel+M_X^2-m_p^2}\Bigg],%\hspace{4em}
\end{aligned}
\end{equation}
\end{widetext}
with
\begin{displaymath}
\qsinel=\frac{\veckt^2+\xi(M_X^2-m_p^2)+\xi^2 m_p^2}{1-\xi}
\end{displaymath}
the photon virtuality after its inelastic emission from the beam particle, and $\xbj = {\qsinel}/({\qsinel+M_X^2-m_p^2})$ the Bjorken scaling variable.
This inelastic density is hence modeling-dependent through the definition of its $F_{2,L}(\xbj,Q^2)$ proton structure functions, along with their linear combination $F_1(\xbj,Q^2)$.
The latter is conventionally defined as:
\begin{displaymath}
\begin{aligned}
  F_1(\xbj,Q^2) = \frac{1}{2\xbj} \bigg[ \left(1+\frac{4m_p^2\xbj^2}{Q^2}\right)F_2(\xbj,Q^2) -{}\\
  {}- F_L(\xbj,Q^2) \bigg].
\end{aligned}
\end{displaymath}

A non-exhaustive list of parametrization implemented in this simulation tool can be found in Section \ref{sect:psf}.

For completeness, the EPA photon fluxes in the collinear approximation commonly used in other simulation tools for photon-induced processes can be obtained by integrating the Budnev fluxes defined in \eqref{eq:budnev_el} and \eqref{eq:budnev_inel} over the full transverse virtuality range of interest, i.e.
\begin{displaymath}
  n^{\rm el,inel}_{\gamma}(\xi) = \int\frac{{\rm d}^2\veckt^2}{\pi\veckt^2} F_{\rm el,inel}^{\gamma}(\xi,\veckt^2).
\end{displaymath}

More generally, it is up to the process developer to add its own modeling of this \kt-parameterized flux, including for other intermediate particles exchanges.
The earlier photon fluxes definitions are however provided as a core features of CepGen.

\section{Implementation}
\label{sect:implem}

For the sake of simplicity, this event simulation tool may be factorized into several independent building blocks.
Its central part is the implementation of the matrix element, as a method computing the scalar event weight for each point computed in the allowed kinematic phase space.
Either the full incoming state is needed as an input to this method, as requested by the original LPAIR code, or the already simplified set of kinematic information needed in the \kt-factorization scheme.

Any process may be defined by the library developer as a class derived from a pure virtual, generic process object provided by the generator core.
It is required to override the three following members:
\begin{itemize}
\item a tool associating the size of the phase space for each of the modes considered (e.g. elastic, single-dissociative, double-dissociative final states for $pp$ processes), and propagated up to the integration algorithm ;
\item a method computing and returning a scalar event weight for any physical point in the phase space, and null elsewhere ;
\item an event definition function associating the kinematic variables of all particles in the event for a given point.
\end{itemize}

This latter populates an internal event object already containing the full incoming state with the central and forward outgoing system for each point.
The user may hence retrieve this object at any stage of the computation process and/or apply a more advanced set of physical selections to fit e.g. a given set of experimental constraints.

To ease the integration of new \kt-factorized process, an additional derivative object provides the full incoming partons (photons, gluons, \ldots) kinematic variables as an input.
Therefore, with the help of a toolbox provided for the definition of inner variables and the Jacobian computation, the processes developer only needs to define a central factorized matrix element and the event structure to allow the process to be fully handled by CepGen.

Furthermore, a FORTRAN interfacing module\footnote{It requires the developer to feed a minimal set of kinematic quantities and event content common blocks, and provide a matrix element definition subroutine.} has been developed to ease the definition of any central \kt-factorized process for a larger fraction of the community.

\subsection{Usage and configuration}\label{sect:usage-config}

Each cross section computation or event generation can be parameterized through a configuration file to be parsed and fed to the internal parameters.
Currently, two formats may be parsed by two handler objects:
\begin{itemize}
\item the \emph{standard LPAIR} steering cards format, combining a four-letters keyword with a configurable value (either an integer, a floating point value, or a characters string).
  It is designed to be fully compatible with all cards previously issued and used in the original version of LPAIR \cite{Baranov:1991yq} ;
\item a structured Python configuration file, defining all parameters collections and their normal usage.
  With the native Python grammar and syntax verification, it allows an easier interaction with the end-user and process developer.
  Additionally, its nested structure enables an easy import of external configuration parameters while reducing the cards verbosity.
\end{itemize}

An example of a comparison between the two steering cards formats for the same run requirements may be found in the supporting website (see Appendix).

\subsection{Proton form factors and structure functions}
\label{sect:psf}

As already introduced, in central exclusive processes, the elastic and semi-elastic scattering of an intermediate particle from the incoming protons may be modeled through its electromagnetic form factors.
Our implementation uses the dipole approximation of the $Q^2$ dependence of the electric and magnetic components introduced in Section \ref{sect:matrixelement-pptoll}.

This elastic scattering of photons from the proton is usually characterized by its very low momentum transfer (or virtuality).
In particular, a dominant fraction of their momentum is emitted collinear to the initial beam.
~\\

With increasing photon virtualities, the probability of protons to leave their bound state to fragment into sub-components increases as well.
As discussed, this behavior is mostly dominated by the inner structure of the proton, characterized by its structure functions.

In the high-energy perturbative limit, these structure functions may be defined according to inner partons distributions in a simple leading-order quark-parton model:
\begin{displaymath}
  F_2(\xbj,Q^2) = \xbj\sum_{f} e_{f}^2 \left[q_f(\xbj,Q^2)+\bar q_f(\xbj,Q^2)\right],
\end{displaymath}
with $e_f$ the quark electric charge (in units of $e$), and $q$ ($\bar q$) the parton density function of the quark (anti-quark) flavor.
An interface between CepGen and the LHAPDF \cite{Buckley:2014ana} library is therefore provided for its vast choice of modelings and its supporting tool for the numerical evaluation of these collinear densities.
~\\

\begin{figure}
  \includegraphics[width=.5\textwidth]{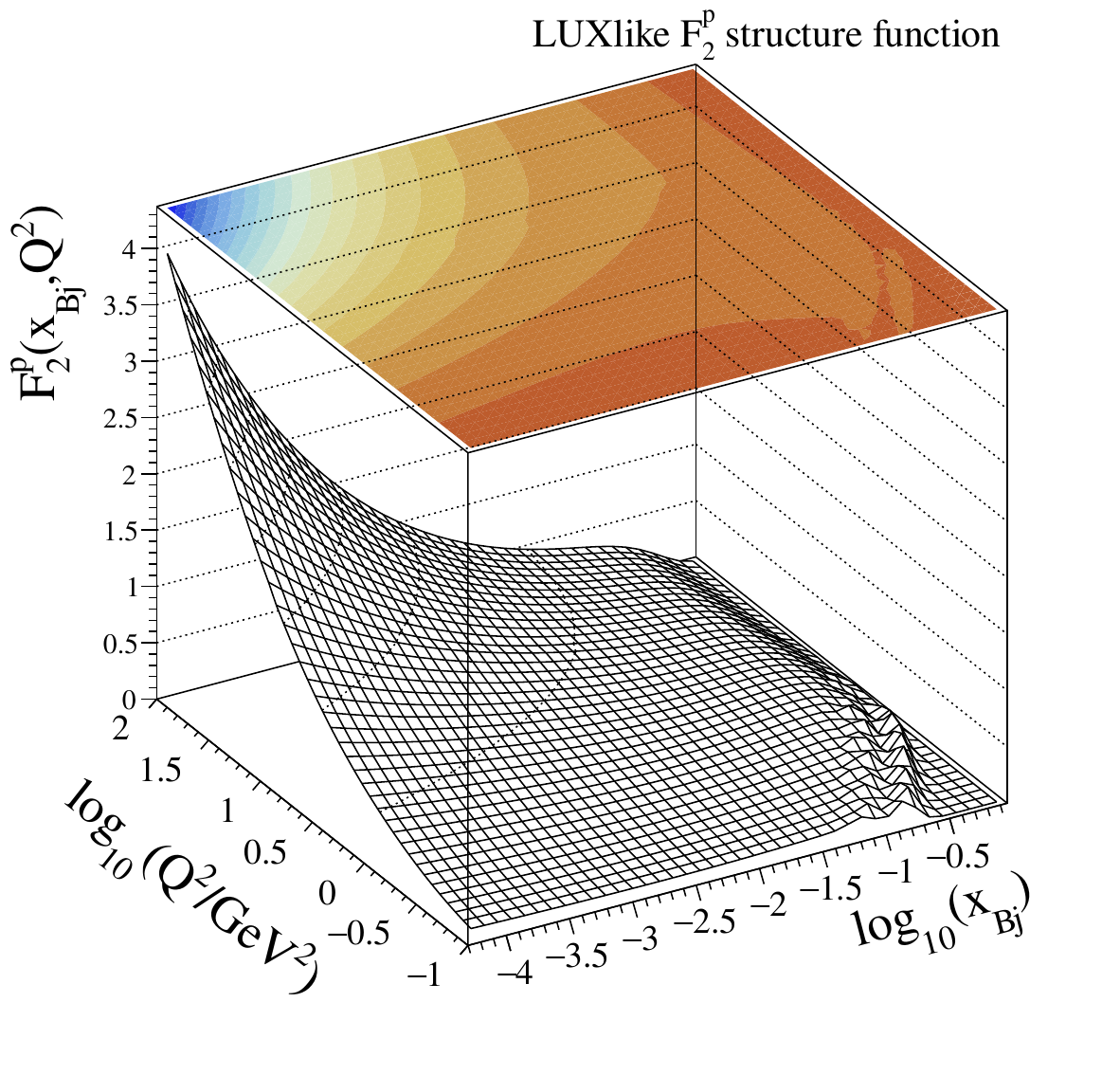}
  \includegraphics[width=.5\textwidth]{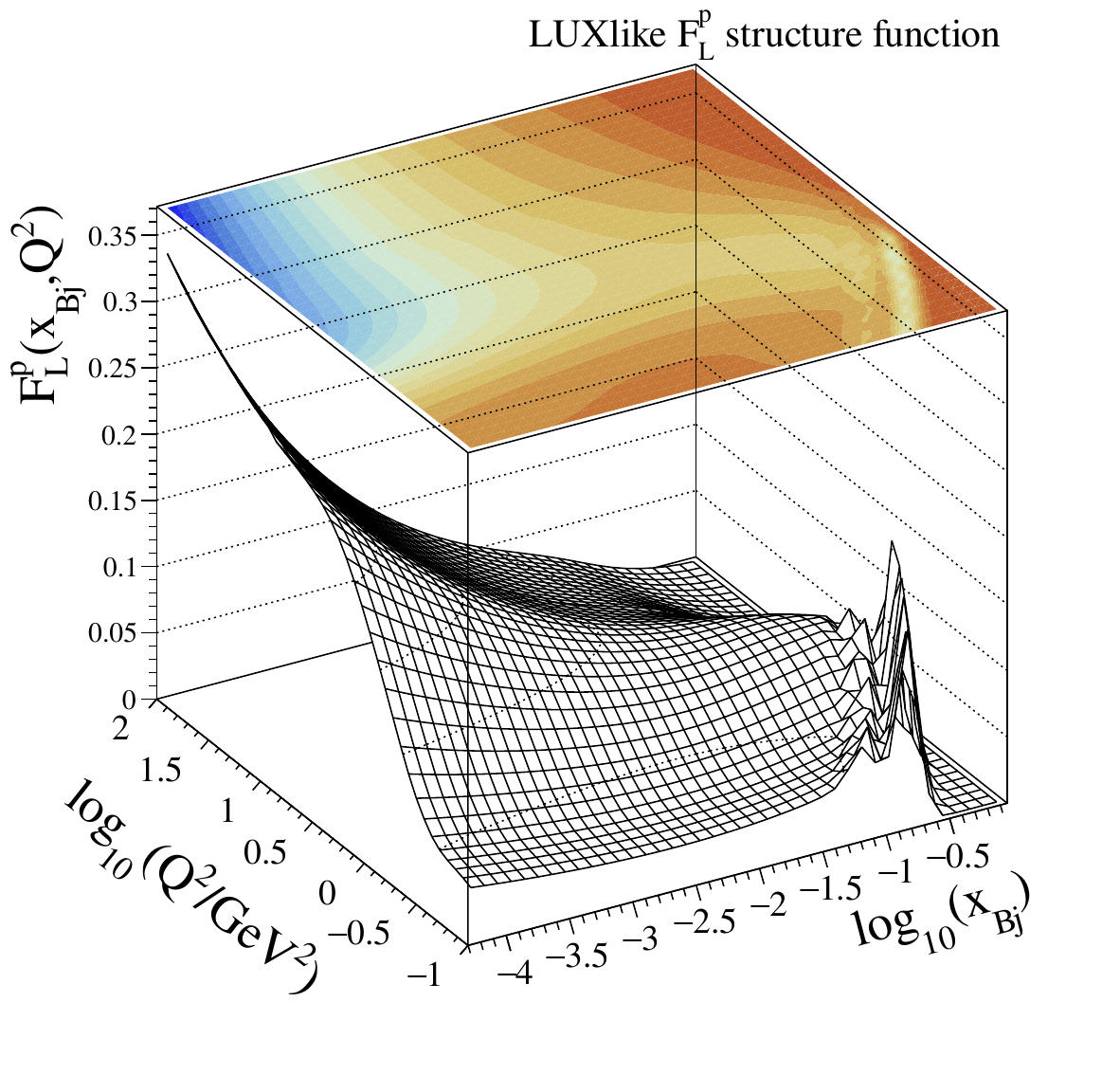}
  \caption{Two-dimensional $(\xbj,Q^2)$ mapping of LUXlike $F_{2/L}$ structure functions defined in the text.
    The perturbative (resonances) region may be found on the left-(right-)most part of both distributions.
    CTEQ6 leading order valence and sea quark distributions are used for the perturbative region.}
  \label{fig:luxlike-sf}
\end{figure}

Additionally, the following parametrizations of $F_{2,L}$ structure functions covering a wide range of fractional momentum losses and photon virtualities are implemented in CepGen:
\begin{itemize}
  \item the Suri-Yennie \cite{Suri:1971yx} parametrization, extracted from the experimental observations at low-$Q^2$ of the total $\gamma p$ cross section in several resonances regions, with invariant masses extending from $1.11$ to $18.03~\gev$ ;
  \item the Fiore {\em et al.} \cite{Fiore:2002re} modeling of the very low $Q^2$ region from the JLab and SLAC observations of resonances, already implemented in LPAIR and reimplemented in CepGen for historical reason ;
  \item the Christy--Bosted \cite{Bosted:2007xd} low-$Q^2$ ($0<Q^2<10~\gev^2$) and low-diffractive mass ($1.1<M_X<3.2~\gev$) modeling, also tuned around the low-mass resonances region ;
  \item the Szczurek--Uleshchenko parametrization \cite{Szczurek:1999rd}, valid at low and intermediate $Q^2$ (up to $100~\gev^2$), and intermediate-$x$ values.
   It relies on a shifted value of the factorization scale ensuring the structure functions convergence at $Q^2\approx 0$ ;
  \item the Abramowicz--Levin--Levy--Maor (ALLM) parametrization \cite{Abramowicz:1991xz,Abramowicz:1997ms} of the intermediate, continuum regime.
   Updated fits containing additional HERMES data samples (GD07p/GD11p) \cite{Gabbert:2007bj,Airapetian:2011nu} are also provided ;
  \item a $F_{2/L}$ grid\footnote{Two-dimensional splines interpolation of a prior grid generation are used for practical reasons, being an extremely time-consuming operation to reproduce for every pair of $(\xbj,Q^2)$.} built from the MSTW partonic density functions \cite{Martin:2009iq} evaluated at next-to-next-to-leading order.
  \item a hybrid ``LUXlike'' set of structure functions with input from the Christy-Bosted resonance, GD11p/ALLM97 continuum, and any perturbative set described earlier.
   This composite modeling, described in details in \cite{Luszczak:2018ntp}, is based on the prescription of \cite{Manohar:2017eqh}.
   A $(\xbj,Q^2)$ mapping of the both structures functions for this set is pictured in Fig.~\ref{fig:luxlike-sf}.
\end{itemize}

\begin{figure*}
  \centering
  \includegraphics[width=.49\textwidth]{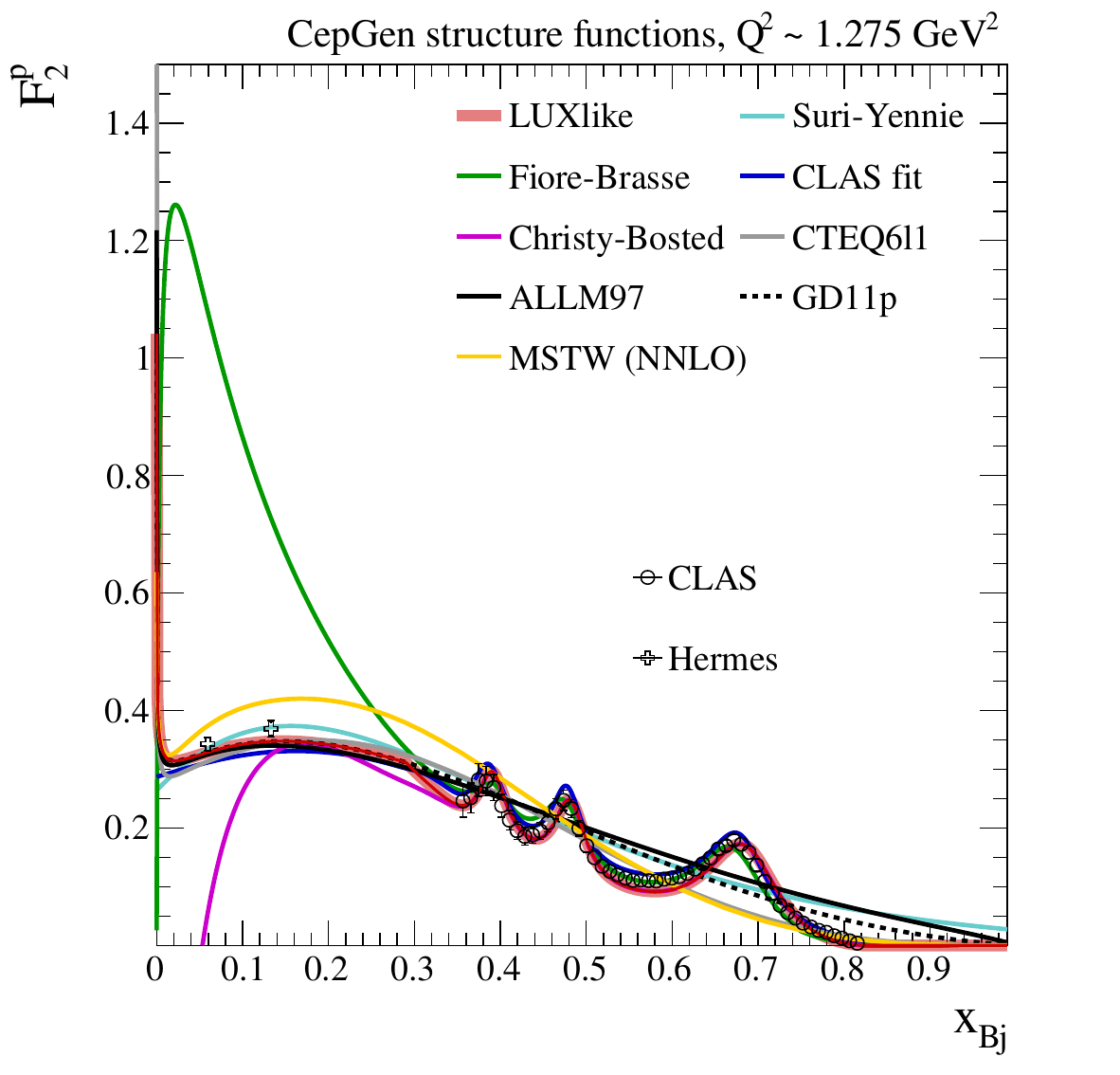}
  \includegraphics[width=.49\textwidth]{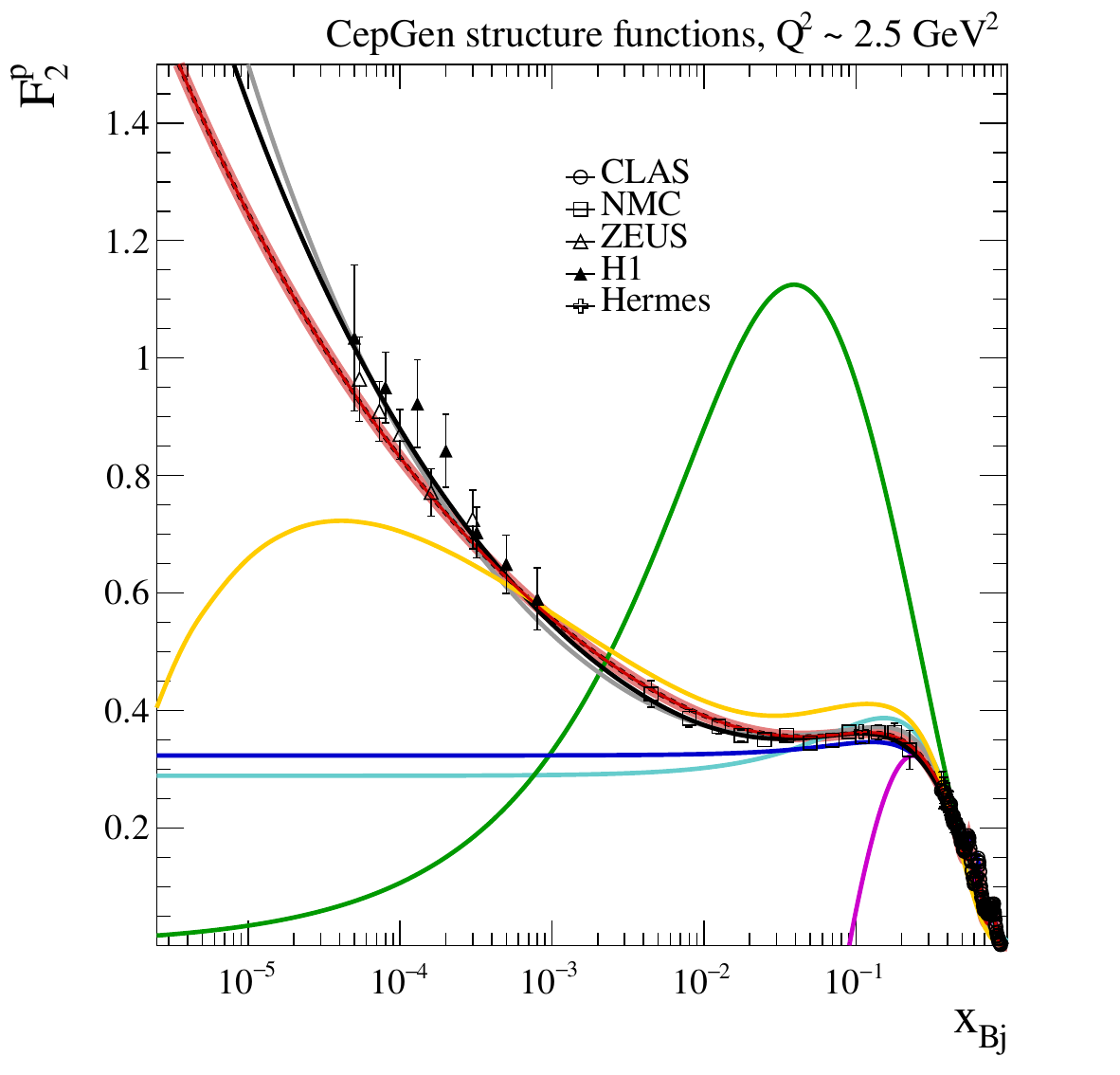}\\
  \includegraphics[width=.49\textwidth]{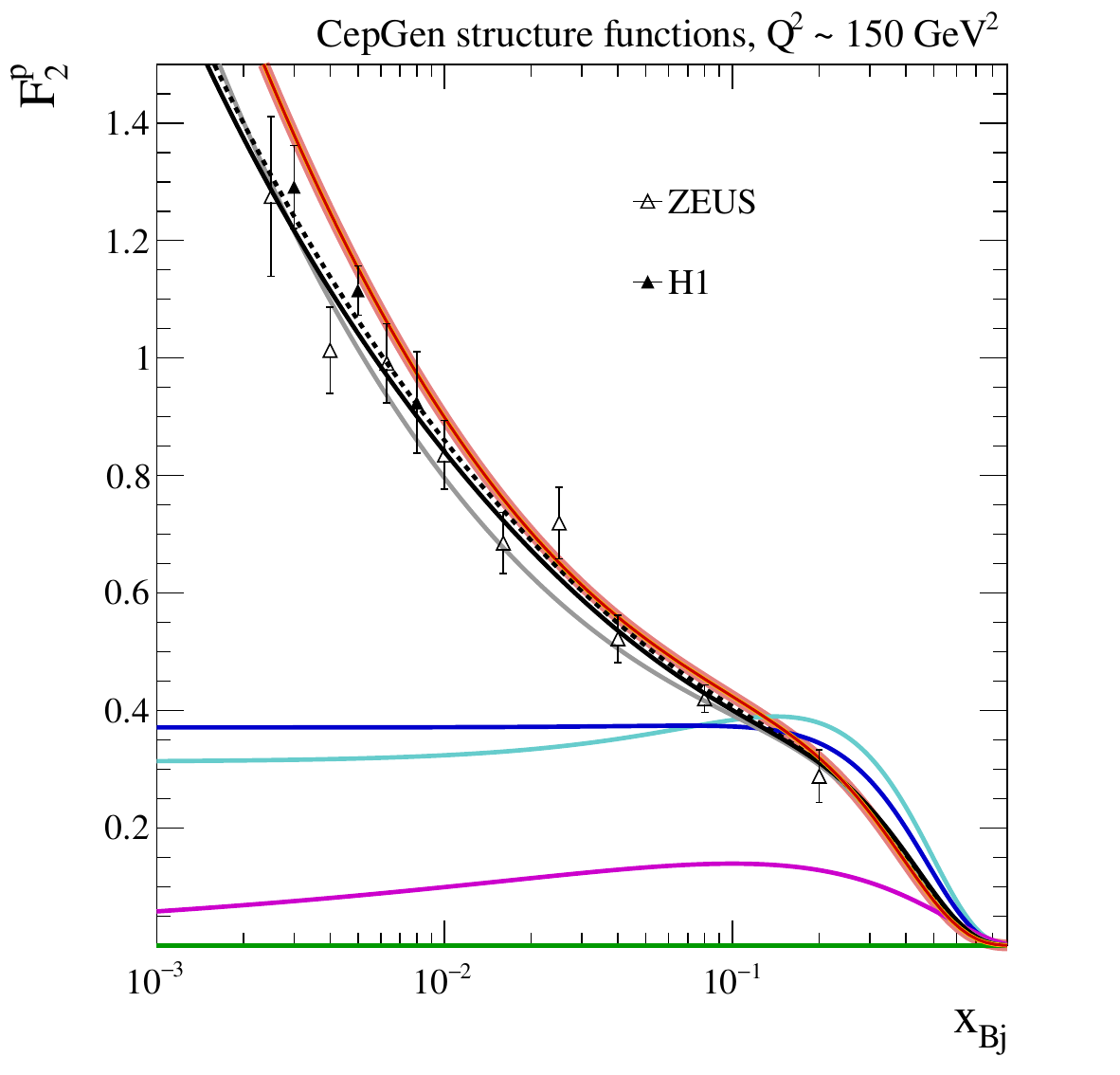}
  \includegraphics[width=.49\textwidth]{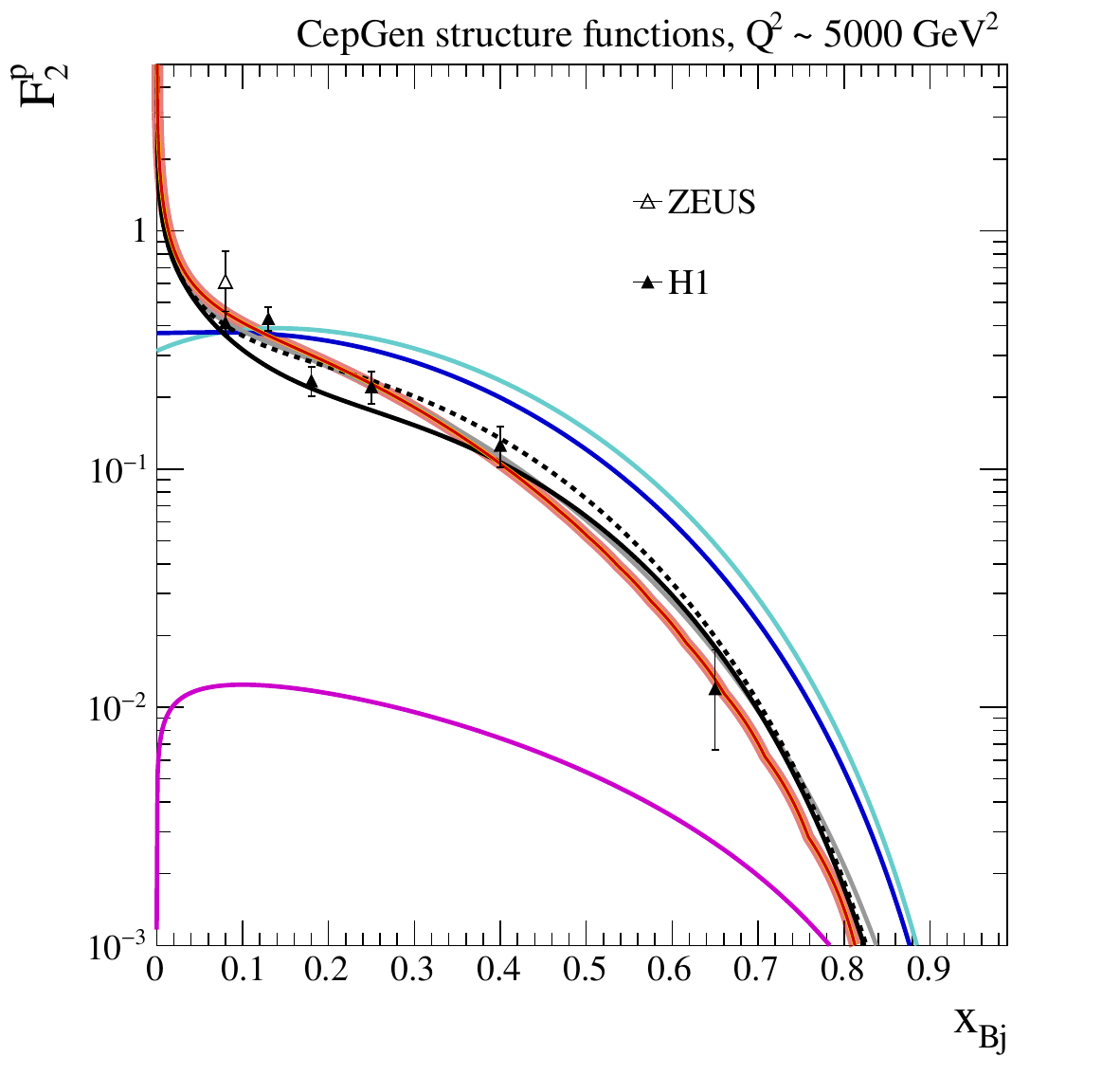}
  \flushleft
  \caption{
    Comparison of the \xbj-dependence of main $F_2$ proton structure functions implemented in CepGen for different $Q^2$ scales.
    Experimental data points shown for comparison are collected from \cite{Benvenuti:1989rh,Gehrmann:1999xn,Airapetian:2011nu}.
    A tolerance of 1\% is set to the $Q^2$ value to account for slight differences in experimental kinematic ranges.
  }
  \label{fig:f2_modellings_comparison}
\end{figure*}

A graphical summary of a good fraction of $F_2$ parametrizations implemented in CepGen can be found in Fig.~\ref{fig:f2_modellings_comparison} for several $Q^2$ momentum scales, along with a subset of experimental observations used in the global fits.
While a fraction of parametrizations is only maintained for historical reasons and ensure backward compatibility with major LPAIR versions, the Suri-Yennie and LUXlike implementations are considered standard for the two-photon production of lepton, gauge boson pairs respectively.

~\\

Depending of the modeling, the longitudinal structure function $F_L$ can either be extracted from the overall fit, or derived from $F_2$ and the longitudinal-to-transverse cross sections ratio $R(\xbj,Q^2)=\sigma_L/\sigma_T$ through the relation:
\begin{displaymath}
  F_L(\xbj,Q^2) = \left(1+\frac{4m_p^2\xbj^2}{Q^2}\right) \frac{R}{1+R}~ F_2(\xbj,Q^2).
\end{displaymath}

Again, a collection of data-driven modelings of this ratio is provided natively within the CepGen structure functions library.
In this first version of the code, the following are implemented:

\begin{itemize}
  \item the E143 fit \cite{Abe:1998ym}, valid for small-\xbj values ($0.03<\xbj<0.1$) ;
  \item the R1990 fit \cite{Whitlow:1990gk}, extracted in the SLAC energy range ($0.6<Q^2<20~{\rm GeV}^2$, $0.1<\xbj<0.9$), and known to give its best predictions for $Q^2>0.3~{\rm GeV}^2$ ;
  %\item the HERMES fit \cite{Airapetian:2011nu}
  \item the Sibirtsev-Blunden parametrization \cite{Sibirtsev:2013cga}, combining measurements from JLab Hall C and SLAC experiments at intermediate $Q^2$.
\end{itemize}
~\\

The full set of structure functions (both $F_2$ and $F_L$, but also the linear combination $F_1$) handled in CepGen can be used in any external user application, being self-contained in a standalone library.
Furthermore, a FORTRAN interface is also provided to ease the user interaction with this collection.

\subsection{Matrix element integration}
\label{sect:matrix-elem-int}
The integration over the full phase space definition is performed using the GSL \cite{Gough:2009:GSL:1538674} implementation of the plain, Vegas \cite{Lepage:1977sw}, and MISER \cite{Press:1989vk} Monte-Carlo integration algorithms.
The first simply probes randomly the full phase space in a fast but inefficient way, while the latter two rely on an optimized choice of the integration grid.

In Vegas, a first probe allows to extract a grid handling the information on all potential numerical singularities.
This operation is critical to ensure the numerical stability of the whole procedure.
In the particular case of two-photon induced processes, several peaking distributions are usually expected (for instance, the photon virtuality $Q^2$ at low values in the case of $t$-channel exchanges).

In a second iteration, the integrand is integrated over the whole phase space, following a generation density determined by this first preparation stage.
The resulting total cross section can hence be estimated within the level of a few percents of precision within only a few iterations.
By default in CepGen, the iteration process is stopped once the $\chi^2$ value is compatible with unity.

In the MISER algorithm, the recursive stratified sampling allows to optimize the phase space point generation through a prior estimate of the higher variance regions to be specifically probed.
A precise estimate of the overall cross section with a lower multiplicity of points probed can therefore rapidly be obtained.

\subsection{Events generation}

The generation of unweighted events to be used as observable states in a simulated process is performed through the \emph{hit or miss} Monte Carlo technique.

For the Vegas integration algorithm, the grid-preparation stage may be used again to optimize the unweighting of events through importance sampling, and reduce the per-phase space point generation time.
This procedure, known as \emph{integrand treatment} may be steered directly by the user.

Since its version 1.0, CepGen introduces an experimental multi-thread implementation of the event generation component.
With this approach, the reduced information sharing between each single thread is allowing for them to build unweighted events in a fast optimized way.
It allows a noticeable reduction of the per-event generation time of most complex processes currently supported.

\subsection{Validation with LPAIR}

\begin{figure}
  \includegraphics[width=.3975\textwidth]{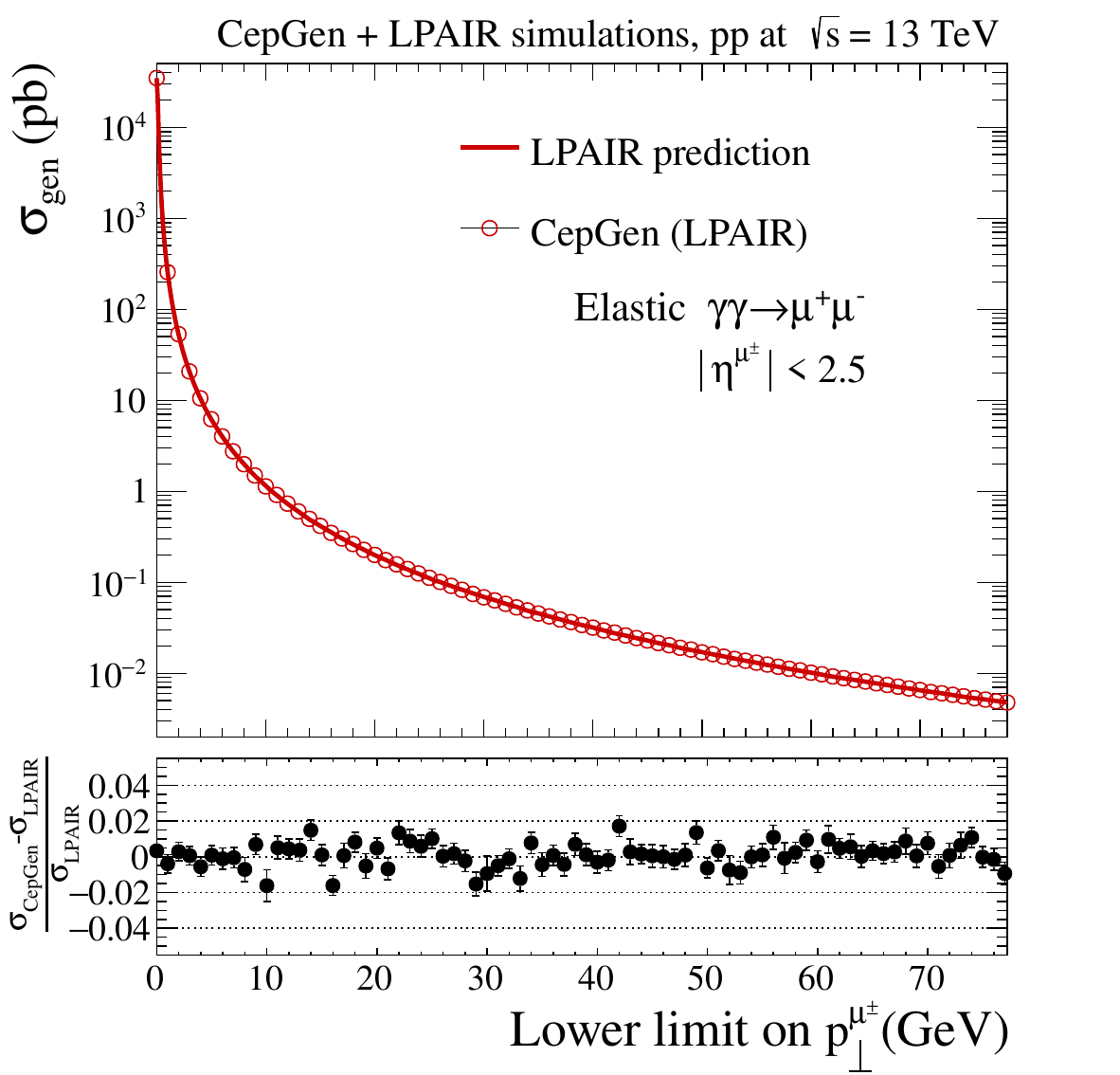}
  \includegraphics[width=.3975\textwidth]{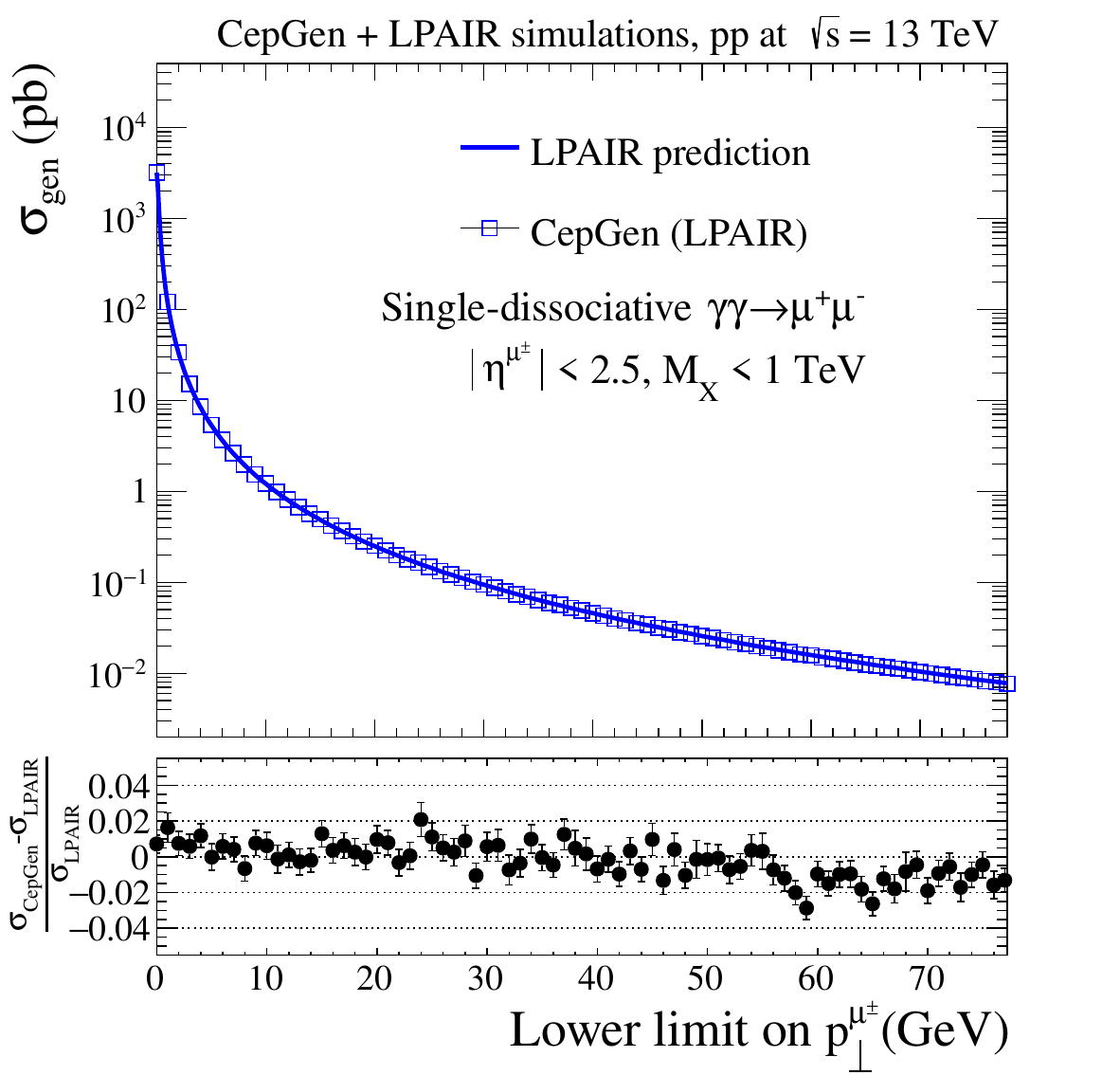}
  \includegraphics[width=.3975\textwidth]{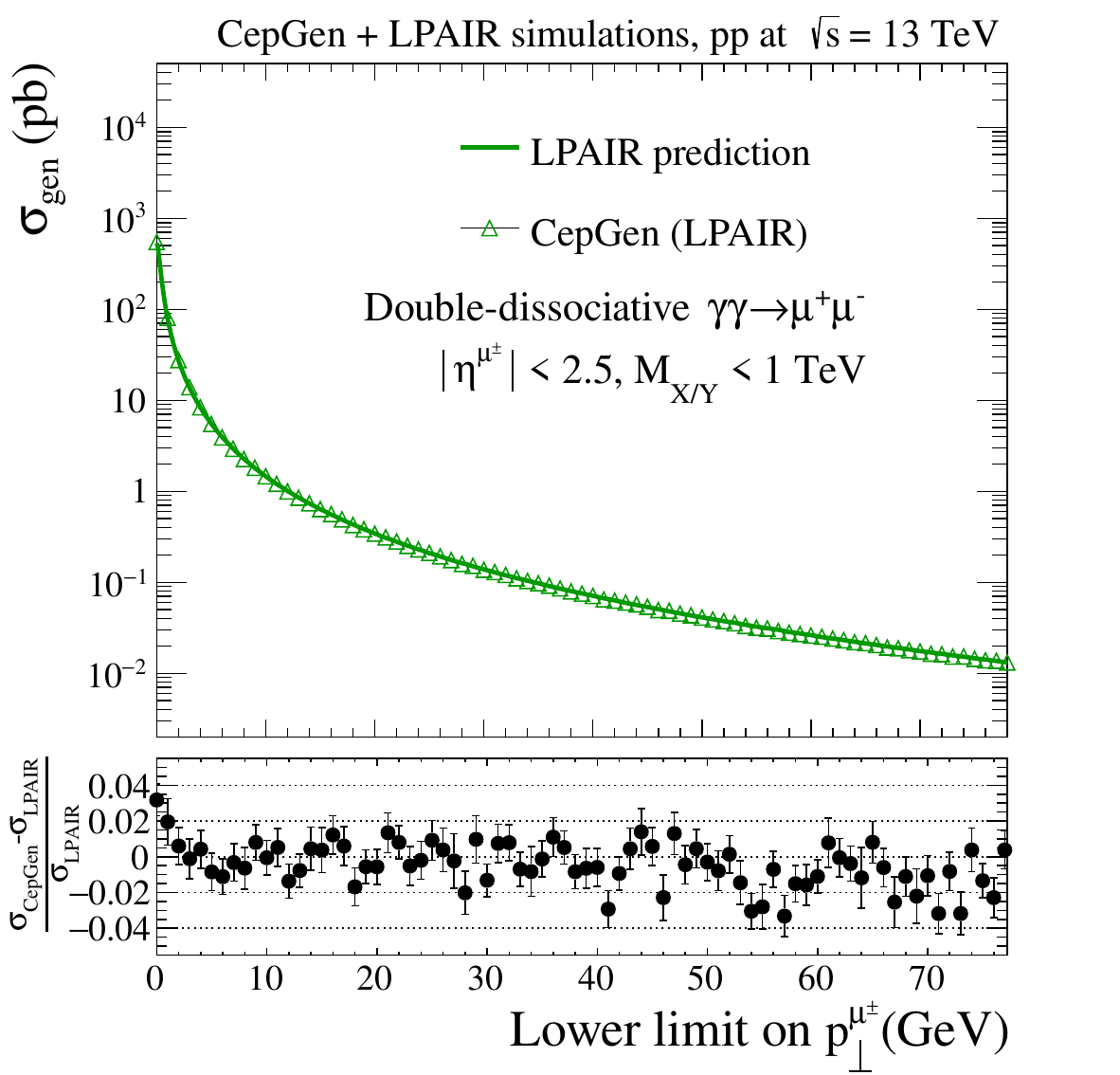}
  \caption{
    Comparison of the elastic, single- and double-dissociative integrated cross sections as computed by LPAIR (solid line), and CepGen (open markers), as a function of the lower transverse momentum selection applied on the outgoing lepton for the $pp \rightarrow p^{(\ast)}(\gamma\gamma\to\mu^+\mu^-)p^{(\ast)}$ process at $\sqrt{s} = 13~{\rm TeV}$.
    The lower part of each distribution pictures the pull distribution for each point scanned.}
  \label{fig:dist_sigma_vs_pt_comp}
\end{figure}

As a validation of the CepGen implementation of the LPAIR \ggll process introduced in Section \ref{sect:matrixelement-lpair}, the generator level cross section can be evaluated for both generators for multiple definitions of the phase space.
In Fig.~\ref{fig:dist_sigma_vs_pt_comp}, the scan of this production cross section as a function of the lower transverse momentum cut applied on both the outgoing leptons is pictured in the three possible proton final states for the original LPAIR algorithm (solid lines) and for this implementation (open markers).

The slight drift in the pull distribution observed at higher values of the single lepton transverse momentum cut are dominantly due to the numerical definition of several physics constants in LPAIR, and corrected in this code.
However, a good agreement can be seen between the two implementations of this matrix element.

\begin{figure}
  \includegraphics[width=.3925\textwidth]{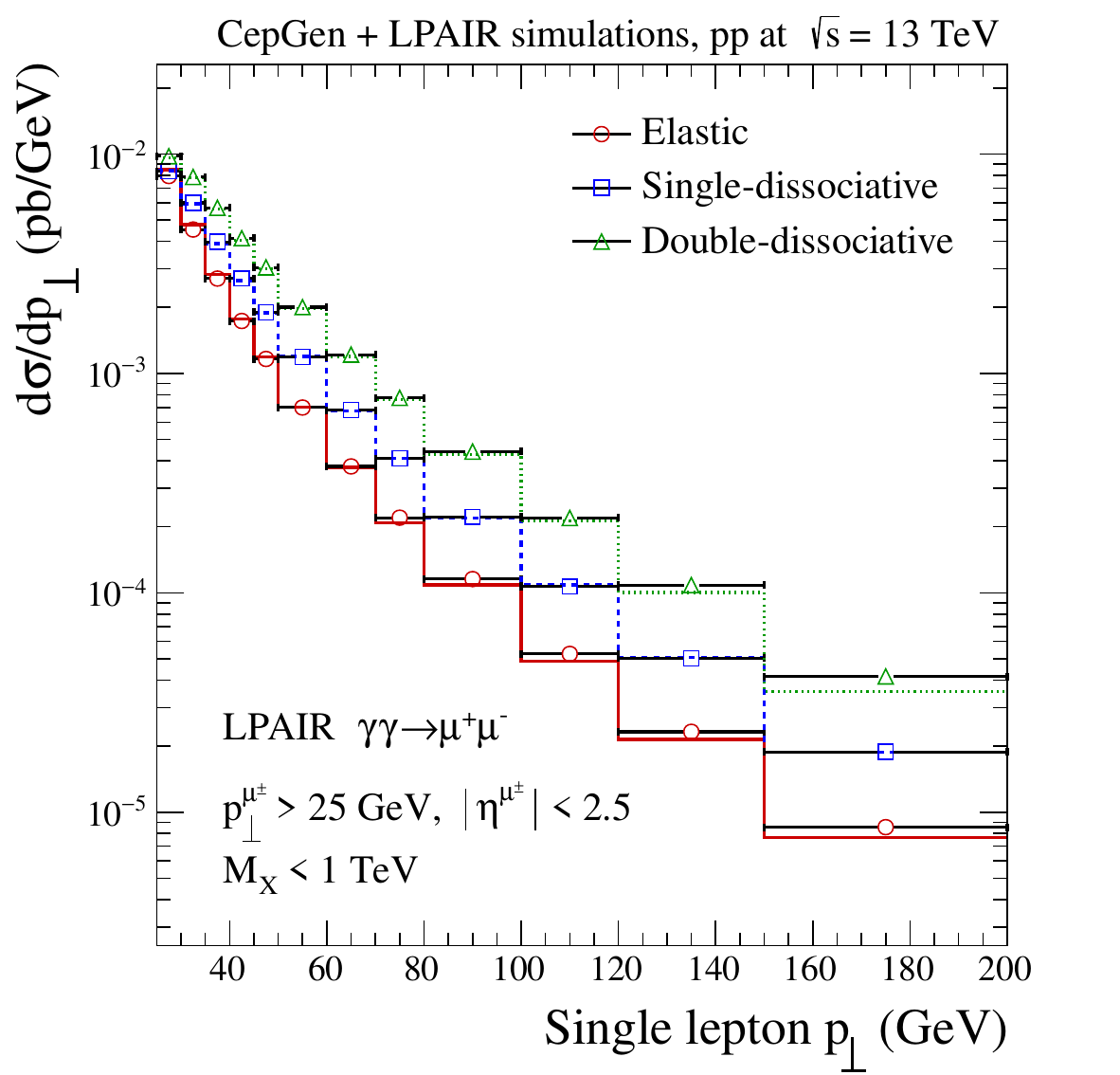}
  \includegraphics[width=.3925\textwidth]{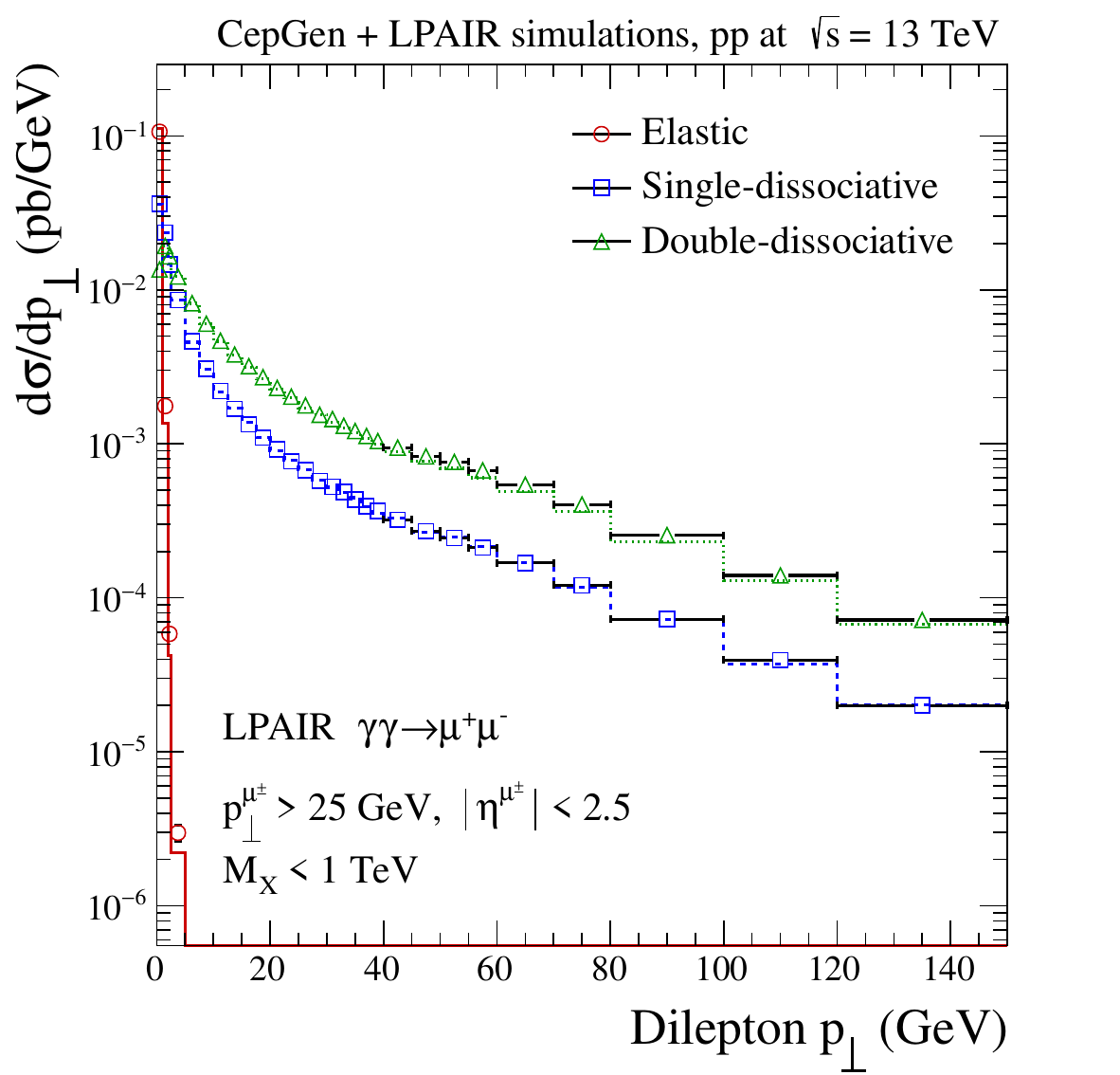}
  \includegraphics[width=.3925\textwidth]{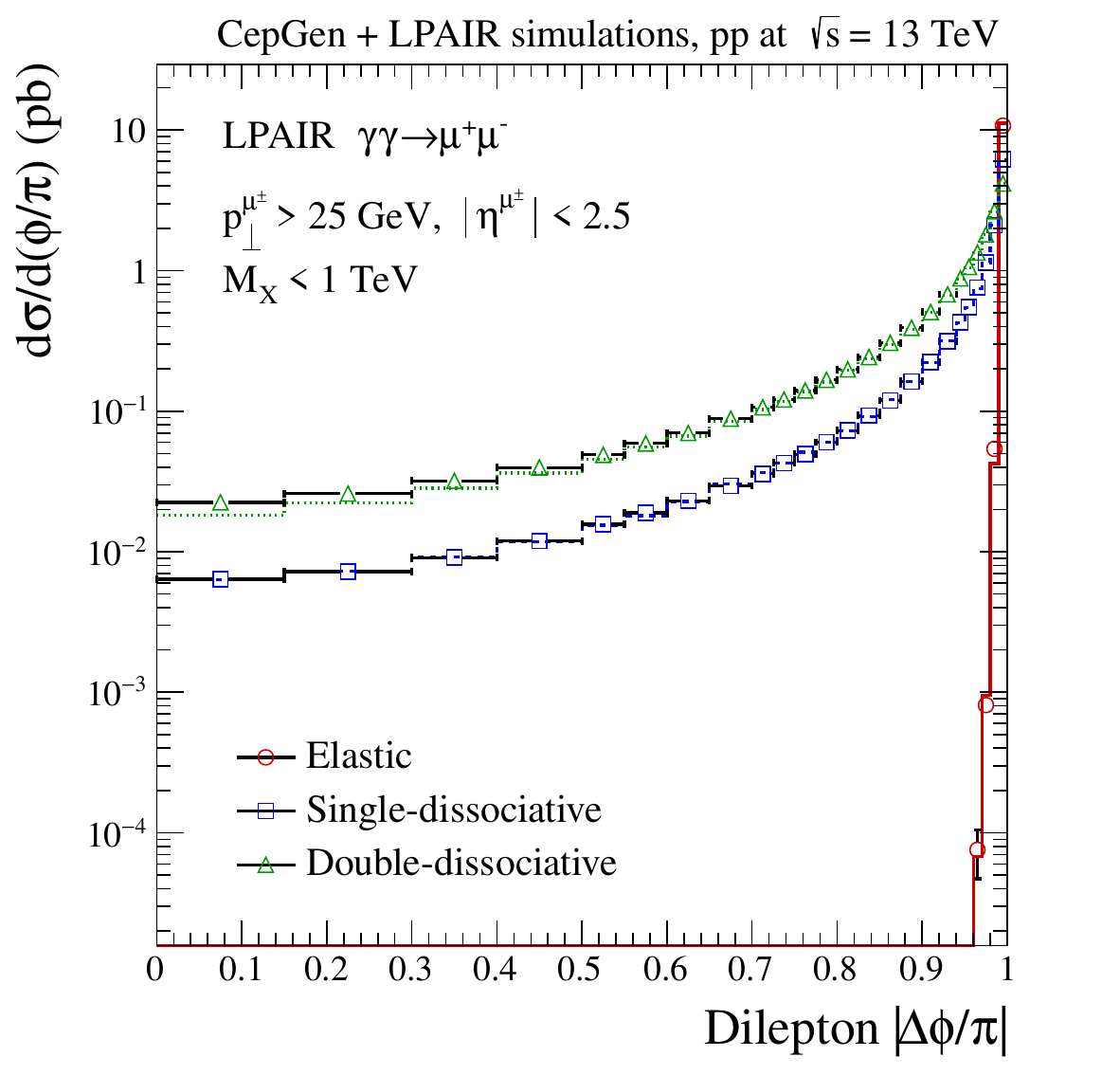}
  \caption{
    Differential cross section distributions for the LPAIR (lines) and CepGen (open markers) implementation of the full matrix element of the $pp\rightarrow p^{(\ast)}(\gamma\gamma\to\mu^+\mu^-)p^{(\ast)}$ process at $\sqrt{s}=13~{\rm TeV}$.
    From top to bottom, single leptons and dilepton transverse momentum, and azimuthal angle difference are pictured.
    Error bars quote the statistical uncertainty on the CepGen estimate for $10^6$ events.}
  \label{fig:dist_lpair_kinematics}
\end{figure}

In Fig.~\ref{fig:dist_lpair_kinematics}, both the CepGen and the LPAIR differential cross section distributions for the dilepton invariant mass, transverse momentum, and the normalized azimuthal correlation between the two leptons are shown, along with the ratio to the LPAIR predictions.
The error bars displayed for all variables account for the statistical uncertainties for the generation of $10^6$ events with CepGen.

Several examples of event generation and cross section computation are provided with the core library to guide the end user towards its integration within the environment.
Multiple wrappers provided along with this software suite allow the direct conversion of the internal event and particles structure to a growing multiplicity of event formats commonly used in high-energy physics\footnote{A few output formats handled in the current version of CepGen are: HepMC ASCII \cite{Dobbs:2001ck}, and \emph{Les Houches Event record} \cite{Alwall:2006yp}.
An example executable also produces simple tree-structured ROOT files \cite{Brun:1997pa} broadly used in the experimental physics community.}.

\section{Additional features}
\label{sect:additions}

Thanks to its modularity, the core features of CepGen can be extended with a set of plugins to ease the end user interaction with external utilities and detector simulation algorithms.

\subsection{Resonances decays and excited proton fragmentation}
\label{sect:ext-hadr}

For this initial release, the PYTHIA 8 \cite{Sjostrand:2014zea} library is interfaced to CepGen to allow its branching fractions, decay products, and decay algorithms implementations to decay all unstable particles.
An additional interfacing module for Herwig 7 \cite{Bahr:2008pv,Bellm:2015jjp} is currently being developed to be released for a later version.

The steering of these external libraries can be performed through the Python cards objects definition.

As observed earlier, the large exchanged photon virtualities may also induce the excitation of the scattered proton leading to its dissociation.
In CepGen, the inelastic photon emission is parameterized through the splitting of this proton into a quark-diquark system.
There, the excited outgoing state is defining the kinematic variables of a valence quark (carrying a fraction \xbj of the outgoing diffractive proton total momentum), thus leaving behind a corresponding color-connected diquark remnant.
This latter is hadronized through the Lund fragmentation algorithm implemented in PYTHIA.

Given its overall knowledge of the full event topology, the multi-parton interactions (MPI), or initial- and final-state radiations frameworks of this latter may be steered to account for larger rapidity gaps suppression in the central detector.
It may hence be adjusted to the drop in survival probabilities observed experimentally in many recent searches \cite{Chatrchyan:2013akv,Khachatryan:2016mud,Aaboud:2017oiq,Cms:2018het}, and lately studied in \cite{Forthomme:2018sxa}.

\subsection{Taming functions}
\label{sect:taming-functions}

For an increased modularity, the differential matrix element can furthermore be modified through a set of taming functions steered by the end user.
It allows to interact directly with the process kinematic variables and account for additional physics effects in the computation of both the differential and integrated cross sections, while leaving behind details of initial parton or hard matrix element modification.

The differential cross section at a given phase space point ${\bf x}$ may hence be modified as:
\begin{displaymath}
  {\rm d}\sigma({\bf x}) \mapsto {\rm d}\sigma'({\bf x}) = \left(\prod_i w_i(x_i)\right)\cdot {\rm d}\sigma({\bf x}),
\end{displaymath}
with $\{w_i\}$ the collection of analytical taming functions to be applied on the resulting matrix element.

Among the possible effects to simulate, one may quote the soft survival probability corrections resulting from large rapidity gap suppression, and observed experimentally in the high-energy limit of central exclusive productions.

This approach has been studied for instance in \cite{Harland-Lang:2016apc}, where a concept of modified photon parton density functions accounting for large rapidity gap suppression is introduced.
The result is an overall modification of the central event kinematic variables.

\begin{figure*}
  \centering
  \includegraphics[width=.49\textwidth]{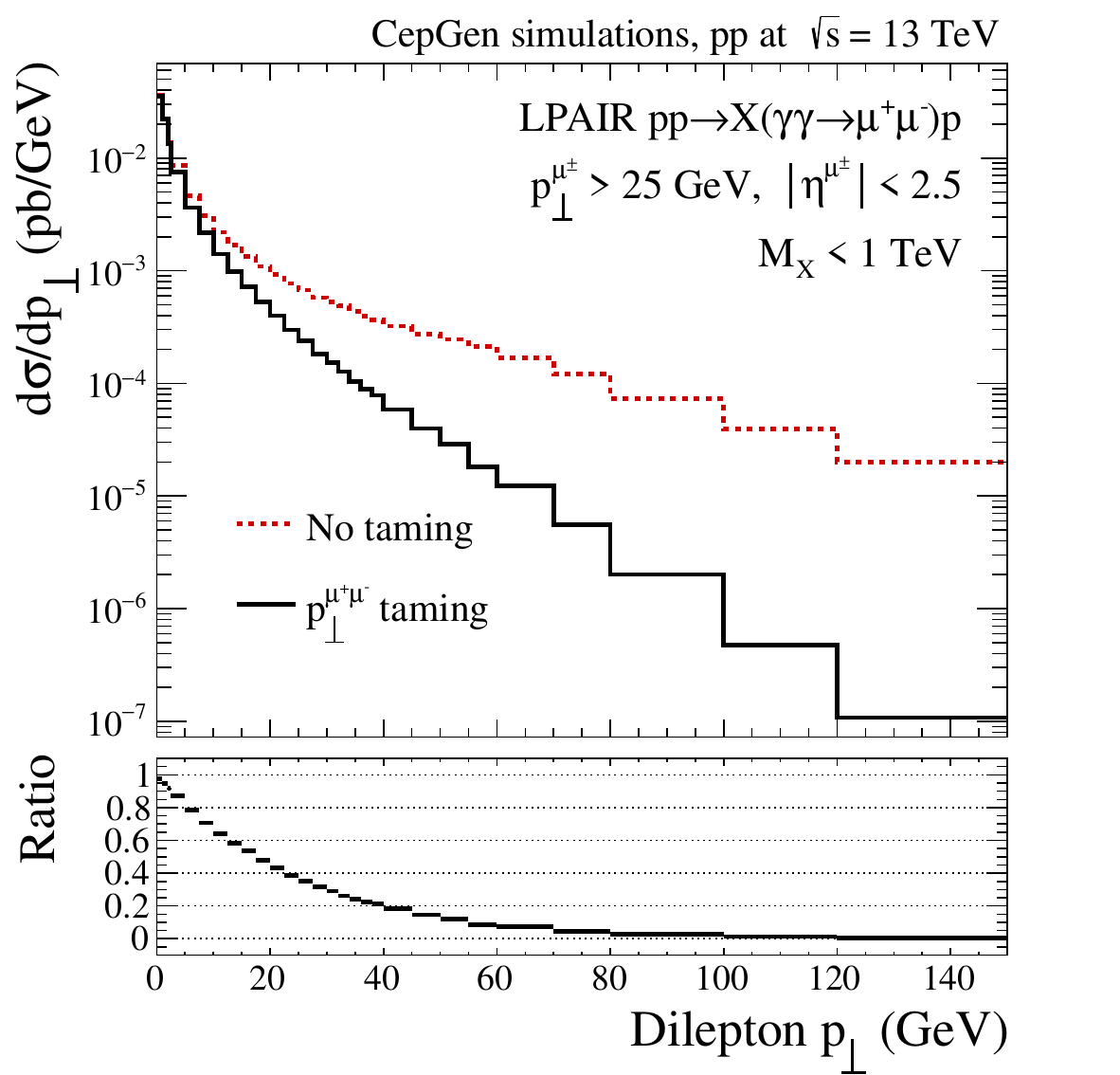}
  \includegraphics[width=.49\textwidth]{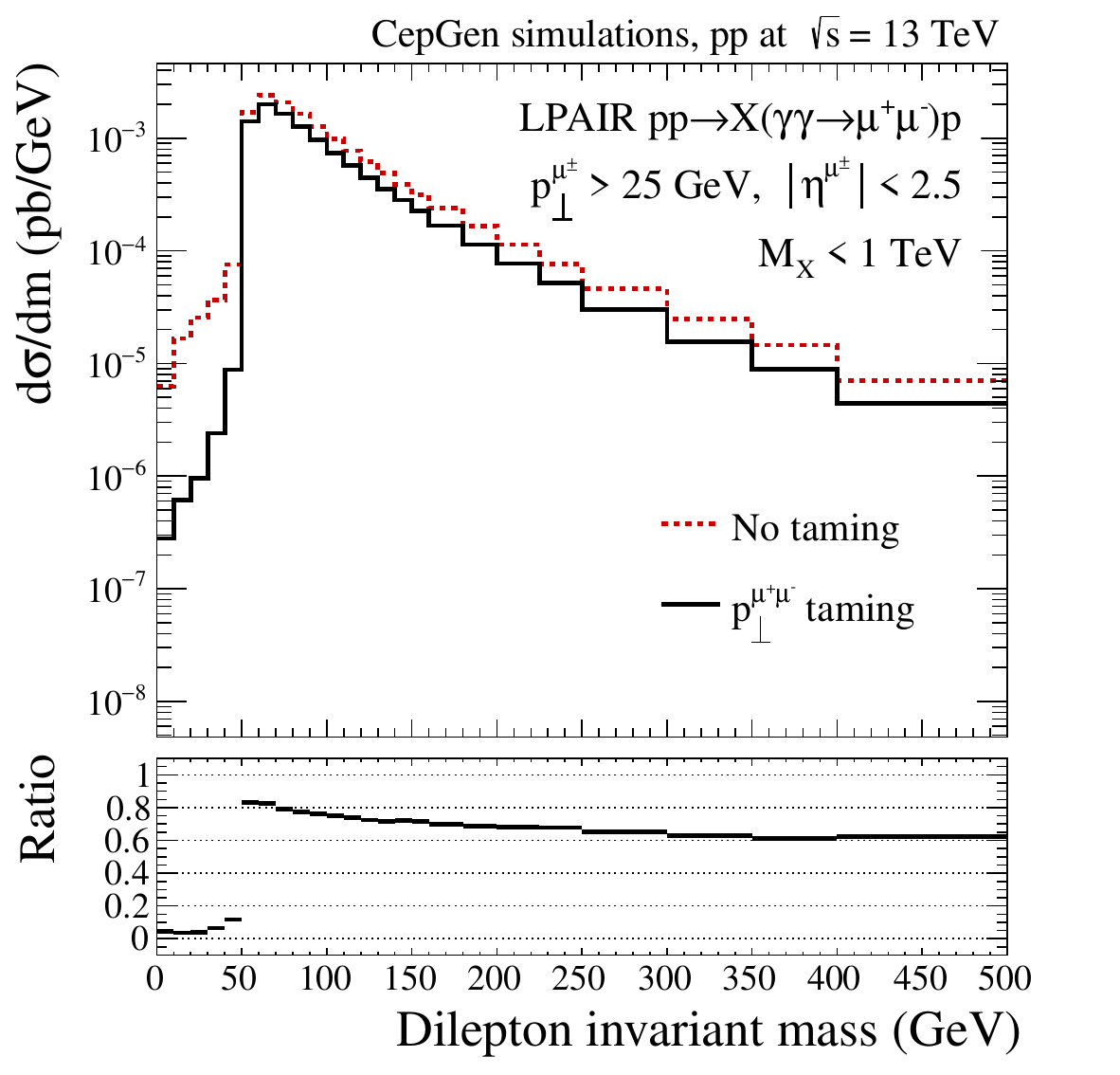}\\
  \includegraphics[width=.49\textwidth]{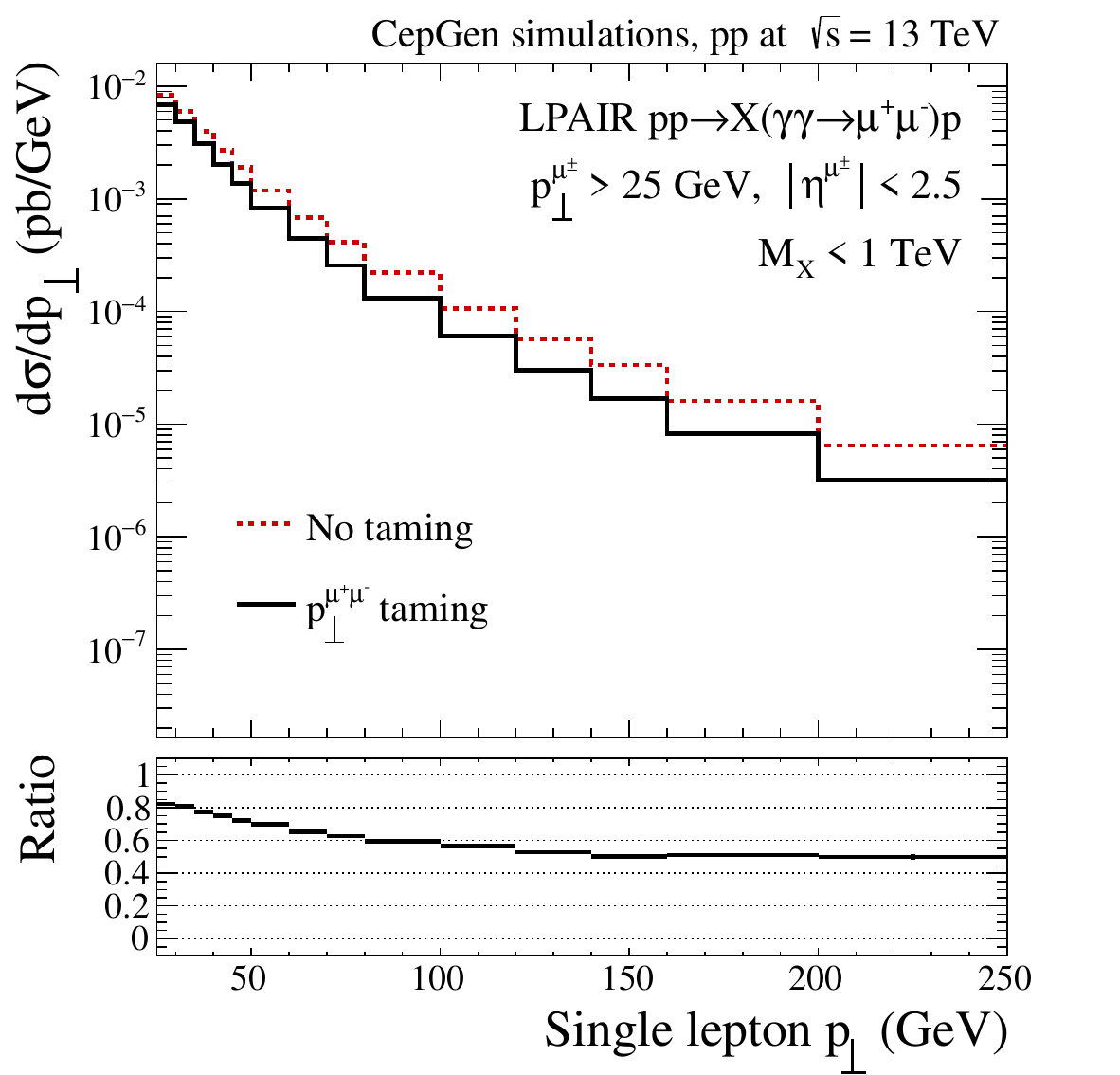}
  \includegraphics[width=.49\textwidth]{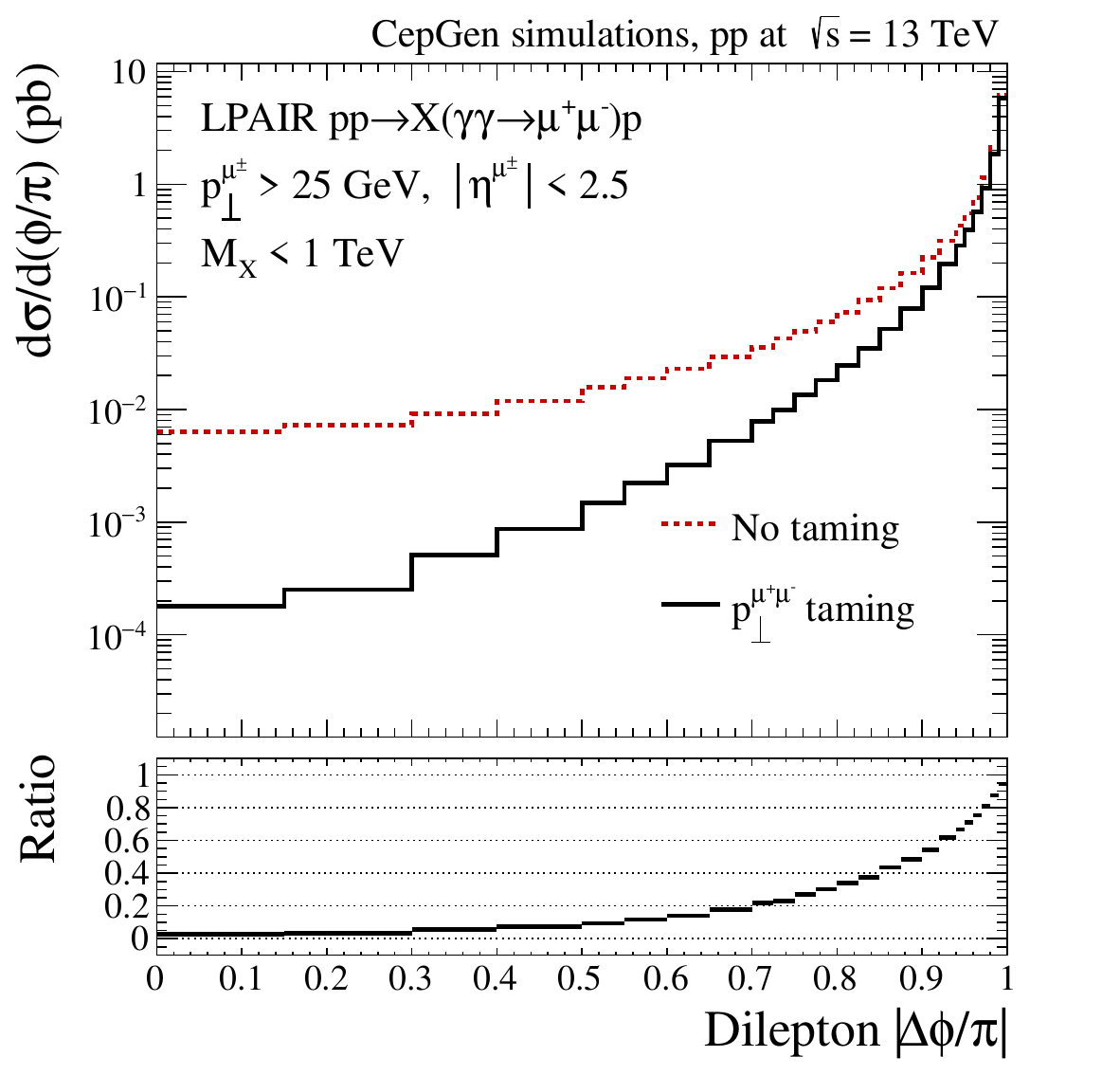}
  \flushleft
  \caption{Effect of introducing a taming function following an exponential suppression in term of the dilepton transverse momentum, as described in the text, in single-dissociative events of the LPAIR process implemented in CepGen.}
  \label{fig:taming_functions}
\end{figure*}

This taming functions framework hence provides a toolbox to evaluate the effects of this complex procedure to the phase space definition.
For instance in the two-photon production of a lepton pair, the rapidity gap survival probability may be treated as a dilepton transverse momentum-dependent correction to follow an exponential reduction, such as:
\begin{displaymath}
  w(\pt^{\mu\mu}) = \exp\left(-0.04\cdot\pt^{\mu\mu}\right).
\end{displaymath}
For the single-dissociative contribution, it corresponds to a survival factor ---after its integration over the full phase space considered--- of $\langle S^2\rangle=0.76$.
The effects of this suppression can be studied in terms of other experimental observables, as pictured in Fig.~\ref{fig:taming_functions}.

\section{Conclusions}
\label{sect:conclusion}

A new framework for the phenomenological and experimental studies of central exclusive processes in the scope of hadron-hadron colliders is presented.
This tool allows both the communities to develop and parameterize any photon- or color singlet-induced process in terms of simple and highly configurable building blocks.

Some of these blocks, as for instance the large collection of structure functions modelings, may also be included individually in any external user library.

Furthermore, both the C++ and  FORTRAN interfaces to processes definition also allows a large fraction of model builders to take part to the development and the study of additional final states, leaving all technicalities of integration, events generation, and external libraries interfacing to be covered by this code.

With an ever increasing number of output formats supported, a direct interfacing to common simulation and reconstruction chains may easily be developed, thus allowing the portage of numerous standalone simulation tools in an embedded framework.

\section*{Acknowledgments}
  The author gratefully acknowledges the support of the Helsinki Institute of Physics for this work.
  He is indebted to M. {\L}uszczak, W. Sch\"{a}fer, A. Szczurek, G. Gil da Silveira, and K. Piotrzkowski for their numerous contributions and feedback through the preparation of this framework.
  He thanks T. Sj\"{o}strand for useful discussions in the PYTHIA 8 linking effort.

\bibliography{main}

%apsrev4-2.bst 2019-01-14 (MD) hand-edited version of apsrev4-1.bst
%Control: key (0)
%Control: author (8) initials jnrlst
%Control: editor formatted (1) identically to author
%Control: production of article title (0) allowed
%Control: page (0) single
%Control: year (1) truncated
%Control: production of eprint (0) enabled
\begin{thebibliography}{40}%
\makeatletter
\providecommand \@ifxundefined [1]{%
 \@ifx{#1\undefined}
}%
\providecommand \@ifnum [1]{%
 \ifnum #1\expandafter \@firstoftwo
 \else \expandafter \@secondoftwo
 \fi
}%
\providecommand \@ifx [1]{%
 \ifx #1\expandafter \@firstoftwo
 \else \expandafter \@secondoftwo
 \fi
}%
\providecommand \natexlab [1]{#1}%
\providecommand \enquote  [1]{``#1''}%
\providecommand \bibnamefont  [1]{#1}%
\providecommand \bibfnamefont [1]{#1}%
\providecommand \citenamefont [1]{#1}%
\providecommand \href@noop [0]{\@secondoftwo}%
\providecommand \href [0]{\begingroup \@sanitize@url \@href}%
\providecommand \@href[1]{\@@startlink{#1}\@@href}%
\providecommand \@@href[1]{\endgroup#1\@@endlink}%
\providecommand \@sanitize@url [0]{\catcode `\\12\catcode `\$12\catcode
  `\&12\catcode `\#12\catcode `\^12\catcode `\_12\catcode `\%12\relax}%
\providecommand \@@startlink[1]{}%
\providecommand \@@endlink[0]{}%
\providecommand \url  [0]{\begingroup\@sanitize@url \@url }%
\providecommand \@url [1]{\endgroup\@href {#1}{\urlprefix }}%
\providecommand \urlprefix  [0]{URL }%
\providecommand \Eprint [0]{\href }%
\providecommand \doibase [0]{https://doi.org/}%
\providecommand \selectlanguage [0]{\@gobble}%
\providecommand \bibinfo  [0]{\@secondoftwo}%
\providecommand \bibfield  [0]{\@secondoftwo}%
\providecommand \translation [1]{[#1]}%
\providecommand \BibitemOpen [0]{}%
\providecommand \bibitemStop [0]{}%
\providecommand \bibitemNoStop [0]{.\EOS\space}%
\providecommand \EOS [0]{\spacefactor3000\relax}%
\providecommand \BibitemShut  [1]{\csname bibitem#1\endcsname}%
\let\auto@bib@innerbib\@empty
%</preamble>
\bibitem [{\citenamefont {Budnev}\ \emph {et~al.}(1975)\citenamefont {Budnev},
  \citenamefont {Ginzburg}, \citenamefont {Meledin},\ and\ \citenamefont
  {Serbo}}]{Budnev:1974de}%
  \BibitemOpen
  \bibfield  {author} {\bibinfo {author} {\bibfnamefont {V.~M.}\ \bibnamefont
  {Budnev}}, \bibinfo {author} {\bibfnamefont {I.~F.}\ \bibnamefont
  {Ginzburg}}, \bibinfo {author} {\bibfnamefont {G.~V.}\ \bibnamefont
  {Meledin}},\ and\ \bibinfo {author} {\bibfnamefont {V.~G.}\ \bibnamefont
  {Serbo}},\ }\bibfield  {title} {\bibinfo {title} {The two photon particle
  production mechanism. {P}hysical problems. {A}pplications. {E}quivalent
  photon approximation},\ }\href {https://doi.org/10.1016/0370-1573(75)90009-5}
  {\bibfield  {journal} {\bibinfo  {journal} {Phys. Rept.}\ }\textbf {\bibinfo
  {volume} {15}},\ \bibinfo {pages} {181} (\bibinfo {year} {1975})}\BibitemShut
  {NoStop}%
%%CITATION = PRPLC,15,181;%%
\bibitem [{\citenamefont {Chatrchyan}\ \emph {et~al.}(2013)\citenamefont
  {Chatrchyan} \emph {et~al.}}]{Chatrchyan:2013akv}%
  \BibitemOpen
  \bibfield  {author} {\bibinfo {author} {\bibfnamefont {S.}~\bibnamefont
  {Chatrchyan}} \emph {et~al.} (\bibinfo {collaboration} {CMS}),\ }\bibfield
  {title} {\bibinfo {title} {Study of exclusive two-photon production of
  {$W^+W^-$} in {$pp$} collisions at {$\sqrt{s} = 7$} {TeV} and constraints on
  anomalous quartic gauge couplings},\ }\href
  {https://doi.org/10.1007/JHEP07(2013)116} {\bibfield  {journal} {\bibinfo
  {journal} {JHEP}\ }\textbf {\bibinfo {volume} {07}},\ \bibinfo {pages}
  {116}},\ \Eprint {https://arxiv.org/abs/1305.5596} {arXiv:1305.5596 [hep-ex]}
  \BibitemShut {NoStop}%
%%CITATION = ARXIV:1305.5596;%%
\bibitem [{\citenamefont {Khachatryan}\ \emph {et~al.}(2016)\citenamefont
  {Khachatryan} \emph {et~al.}}]{Khachatryan:2016mud}%
  \BibitemOpen
  \bibfield  {author} {\bibinfo {author} {\bibfnamefont {V.}~\bibnamefont
  {Khachatryan}} \emph {et~al.} (\bibinfo {collaboration} {CMS}),\ }\bibfield
  {title} {\bibinfo {title} {Evidence for exclusive {$\gamma\gamma \to W^+
  W^-$} production and constraints on anomalous quartic gauge couplings in
  {$pp$} collisions at {$\sqrt{s}=7$} and 8 {TeV}},\ }\href
  {https://doi.org/10.1007/JHEP08(2016)119} {\bibfield  {journal} {\bibinfo
  {journal} {JHEP}\ }\textbf {\bibinfo {volume} {08}},\ \bibinfo {pages}
  {119}},\ \Eprint {https://arxiv.org/abs/1604.04464} {arXiv:1604.04464
  [hep-ex]} \BibitemShut {NoStop}%
%%CITATION = ARXIV:1604.04464;%%
\bibitem [{\citenamefont {Aaboud}\ \emph {et~al.}(2018)\citenamefont {Aaboud}
  \emph {et~al.}}]{Aaboud:2017oiq}%
  \BibitemOpen
  \bibfield  {author} {\bibinfo {author} {\bibfnamefont {M.}~\bibnamefont
  {Aaboud}} \emph {et~al.} (\bibinfo {collaboration} {ATLAS}),\ }\bibfield
  {title} {\bibinfo {title} {Measurement of the exclusive {$\gamma \gamma
  \rightarrow \mu^+ \mu^-$} process in proton-proton collisions at
  {$\sqrt{s}=13$} {TeV} with the {ATLAS} detector},\ }\href
  {https://doi.org/10.1016/j.physletb.2017.12.043} {\bibfield  {journal}
  {\bibinfo  {journal} {Phys. Lett. B}\ }\textbf {\bibinfo {volume} {777}},\
  \bibinfo {pages} {303} (\bibinfo {year} {2018})},\ \Eprint
  {https://arxiv.org/abs/1708.04053} {arXiv:1708.04053 [hep-ex]} \BibitemShut
  {NoStop}%
%%CITATION = ARXIV:1708.04053;%%
\bibitem [{\citenamefont {{\L{}}uszczak}\ \emph {et~al.}(2018)\citenamefont
  {{\L{}}uszczak}, \citenamefont {Sch{\"{a}}fer},\ and\ \citenamefont
  {Szczurek}}]{Luszczak:2018ntp}%
  \BibitemOpen
  \bibfield  {author} {\bibinfo {author} {\bibfnamefont {M.}~\bibnamefont
  {{\L{}}uszczak}}, \bibinfo {author} {\bibfnamefont {W.}~\bibnamefont
  {Sch{\"{a}}fer}},\ and\ \bibinfo {author} {\bibfnamefont {A.}~\bibnamefont
  {Szczurek}},\ }\bibfield  {title} {\bibinfo {title} {Production of {$W^+
  W^-$} pairs via {$\gamma^\ast\gamma^\ast \to W^+ W^-$} subprocess with photon
  transverse momenta},\ }\href {https://doi.org/10.1007/JHEP05(2018)064}
  {\bibfield  {journal} {\bibinfo  {journal} {JHEP}\ }\textbf {\bibinfo
  {volume} {05}},\ \bibinfo {pages} {064}},\ \Eprint
  {https://arxiv.org/abs/1802.03244} {arXiv:1802.03244 [hep-ph]} \BibitemShut
  {NoStop}%
%%CITATION = ARXIV:1802.03244;%%
\bibitem [{\citenamefont {List}\ and\ \citenamefont
  {Mastroberardino}(1998)}]{List:1998jz}%
  \BibitemOpen
  \bibfield  {author} {\bibinfo {author} {\bibfnamefont {B.}~\bibnamefont
  {List}}\ and\ \bibinfo {author} {\bibfnamefont {A.}~\bibnamefont
  {Mastroberardino}},\ }\bibfield  {title} {\bibinfo {title} {{DIFFVM}: A
  {Monte Carlo} generator for diffractive processes in {ep} scattering},\
  }\bibfield  {booktitle} {\emph {\bibinfo {booktitle} {{Monte Carlo}
  generators for {HERA} physics. Proceedings, Workshop, Hamburg, Germany,
  1998-1999}},\ }\href@noop {} {\bibfield  {journal} {\bibinfo  {journal}
  {Conf. Proc. C}\ }\textbf {\bibinfo {volume} {980427}},\ \bibinfo {pages}
  {396} (\bibinfo {year} {1998})}\BibitemShut {NoStop}%
%%CITATION = CONFP,C980427,396;%%
\bibitem [{\citenamefont {Kimber}\ \emph {et~al.}(2001)\citenamefont {Kimber},
  \citenamefont {Martin},\ and\ \citenamefont {Ryskin}}]{Kimber:2001sc}%
  \BibitemOpen
  \bibfield  {author} {\bibinfo {author} {\bibfnamefont {M.~A.}\ \bibnamefont
  {Kimber}}, \bibinfo {author} {\bibfnamefont {A.~D.}\ \bibnamefont {Martin}},\
  and\ \bibinfo {author} {\bibfnamefont {M.~G.}\ \bibnamefont {Ryskin}},\
  }\bibfield  {title} {\bibinfo {title} {Unintegrated parton distributions},\
  }\href {https://doi.org/10.1103/PhysRevD.63.114027} {\bibfield  {journal}
  {\bibinfo  {journal} {Phys. Rev. D}\ }\textbf {\bibinfo {volume} {63}},\
  \bibinfo {pages} {114027} (\bibinfo {year} {2001})},\ \Eprint
  {https://arxiv.org/abs/hep-ph/0101348} {arXiv:hep-ph/0101348 [hep-ph]}
  \BibitemShut {NoStop}%
%%CITATION = HEP-PH/0101348;%%
\bibitem [{\citenamefont {Watt}\ \emph {et~al.}(2004)\citenamefont {Watt},
  \citenamefont {Martin},\ and\ \citenamefont {Ryskin}}]{Watt:2003vf}%
  \BibitemOpen
  \bibfield  {author} {\bibinfo {author} {\bibfnamefont {G.}~\bibnamefont
  {Watt}}, \bibinfo {author} {\bibfnamefont {A.~D.}\ \bibnamefont {Martin}},\
  and\ \bibinfo {author} {\bibfnamefont {M.~G.}\ \bibnamefont {Ryskin}},\
  }\bibfield  {title} {\bibinfo {title} {Unintegrated parton distributions and
  electroweak boson production at hadron colliders},\ }\href
  {https://doi.org/10.1103/PhysRevD.70.014012, 10.1103/PhysRevD.70.079902}
  {\bibfield  {journal} {\bibinfo  {journal} {Phys. Rev. D}\ }\textbf {\bibinfo
  {volume} {70}},\ \bibinfo {pages} {014012} (\bibinfo {year} {2004})},\
  \bibinfo {note} {[Erratum: Phys. Rev. {\bf D70}, 079902 (2004)]},\ \Eprint
  {https://arxiv.org/abs/hep-ph/0309096} {arXiv:hep-ph/0309096 [hep-ph]}
  \BibitemShut {NoStop}%
%%CITATION = HEP-PH/0309096;%%
\bibitem [{\citenamefont {Vermaseren}(1983)}]{Vermaseren:1982cz}%
  \BibitemOpen
  \bibfield  {author} {\bibinfo {author} {\bibfnamefont {J.~A.~M.}\
  \bibnamefont {Vermaseren}},\ }\bibfield  {title} {\bibinfo {title} {Two
  photon processes at very high-energies},\ }\href
  {https://doi.org/10.1016/0550-3213(83)90336-X} {\bibfield  {journal}
  {\bibinfo  {journal} {Nucl. Phys. B}\ }\textbf {\bibinfo {volume} {229}},\
  \bibinfo {pages} {347} (\bibinfo {year} {1983})}\BibitemShut {NoStop}%
%%CITATION = NUPHA,B229,347;%%
\bibitem [{\citenamefont {Baranov}\ \emph {et~al.}(1991)\citenamefont
  {Baranov}, \citenamefont {Duenger}, \citenamefont {Shooshtari},\ and\
  \citenamefont {Vermaseren}}]{Baranov:1991yq}%
  \BibitemOpen
  \bibfield  {author} {\bibinfo {author} {\bibfnamefont {S.~P.}\ \bibnamefont
  {Baranov}}, \bibinfo {author} {\bibfnamefont {O.}~\bibnamefont {Duenger}},
  \bibinfo {author} {\bibfnamefont {H.}~\bibnamefont {Shooshtari}},\ and\
  \bibinfo {author} {\bibfnamefont {J.~A.~M.}\ \bibnamefont {Vermaseren}},\
  }\bibfield  {title} {\bibinfo {title} {{LPAIR}: A generator for lepton pair
  production},\ }in\ \href@noop {} {\emph {\bibinfo {booktitle} {{Workshop on
  Physics at HERA Hamburg, Germany}}}}\ (\bibinfo {year} {1991})\ p.\ \bibinfo
  {pages} {1478}\BibitemShut {NoStop}%
%%CITATION = INSPIRE-326934;%%
\bibitem [{\citenamefont {Gil~da Silveira}\ \emph {et~al.}(2015)\citenamefont
  {Gil~da Silveira}, \citenamefont {Forthomme}, \citenamefont {Piotrzkowski},
  \citenamefont {Sch{\"{a}}fer},\ and\ \citenamefont
  {Szczurek}}]{daSilveira:2014jla}%
  \BibitemOpen
  \bibfield  {author} {\bibinfo {author} {\bibfnamefont {G.}~\bibnamefont
  {Gil~da Silveira}}, \bibinfo {author} {\bibfnamefont {L.}~\bibnamefont
  {Forthomme}}, \bibinfo {author} {\bibfnamefont {K.}~\bibnamefont
  {Piotrzkowski}}, \bibinfo {author} {\bibfnamefont {W.}~\bibnamefont
  {Sch{\"{a}}fer}},\ and\ \bibinfo {author} {\bibfnamefont {A.}~\bibnamefont
  {Szczurek}},\ }\bibfield  {title} {\bibinfo {title} {Central
  {$\mu^{+}\mu^{-}$} production via photon-photon fusion in proton-proton
  collisions with proton dissociation},\ }\href
  {https://doi.org/10.1007/JHEP02(2015)159} {\bibfield  {journal} {\bibinfo
  {journal} {JHEP}\ }\textbf {\bibinfo {volume} {1502}},\ \bibinfo {pages}
  {159}},\ \Eprint {https://arxiv.org/abs/1409.1541} {arXiv:1409.1541 [hep-ph]}
  \BibitemShut {NoStop}%
%%CITATION = ARXIV:1409.1541;%%
\bibitem [{\citenamefont {{\L{}}uszczak}\ \emph {et~al.}(2016)\citenamefont
  {{\L{}}uszczak}, \citenamefont {Sch{\"{a}}fer},\ and\ \citenamefont
  {Szczurek}}]{Luszczak:2015aoa}%
  \BibitemOpen
  \bibfield  {author} {\bibinfo {author} {\bibfnamefont {M.}~\bibnamefont
  {{\L{}}uszczak}}, \bibinfo {author} {\bibfnamefont {W.}~\bibnamefont
  {Sch{\"{a}}fer}},\ and\ \bibinfo {author} {\bibfnamefont {A.}~\bibnamefont
  {Szczurek}},\ }\bibfield  {title} {\bibinfo {title} {Two-photon dilepton
  production in proton-proton collisions: two alternative approaches},\ }\href
  {https://doi.org/10.1103/PhysRevD.93.074018} {\bibfield  {journal} {\bibinfo
  {journal} {Phys. Rev. D}\ }\textbf {\bibinfo {volume} {93}},\ \bibinfo
  {pages} {074018} (\bibinfo {year} {2016})},\ \Eprint
  {https://arxiv.org/abs/1510.00294} {arXiv:1510.00294 [hep-ph]} \BibitemShut
  {NoStop}%
%%CITATION = ARXIV:1510.00294;%%
\bibitem [{\citenamefont {Buckley}\ \emph {et~al.}(2015)\citenamefont
  {Buckley}, \citenamefont {Ferrando}, \citenamefont {Lloyd}, \citenamefont
  {Nordstr{\"{o}}m}, \citenamefont {Page}, \citenamefont {R{\"{u}}fenacht},
  \citenamefont {Sch{\"{o}}nherr},\ and\ \citenamefont
  {Watt}}]{Buckley:2014ana}%
  \BibitemOpen
  \bibfield  {author} {\bibinfo {author} {\bibfnamefont {A.}~\bibnamefont
  {Buckley}}, \bibinfo {author} {\bibfnamefont {J.}~\bibnamefont {Ferrando}},
  \bibinfo {author} {\bibfnamefont {S.}~\bibnamefont {Lloyd}}, \bibinfo
  {author} {\bibfnamefont {K.}~\bibnamefont {Nordstr{\"{o}}m}}, \bibinfo
  {author} {\bibfnamefont {B.}~\bibnamefont {Page}}, \bibinfo {author}
  {\bibfnamefont {M.}~\bibnamefont {R{\"{u}}fenacht}}, \bibinfo {author}
  {\bibfnamefont {M.}~\bibnamefont {Sch{\"{o}}nherr}},\ and\ \bibinfo {author}
  {\bibfnamefont {G.}~\bibnamefont {Watt}},\ }\bibfield  {title} {\bibinfo
  {title} {{LHAPDF6}: parton density access in the {LHC} precision era},\
  }\href {https://doi.org/10.1140/epjc/s10052-015-3318-8} {\bibfield  {journal}
  {\bibinfo  {journal} {Eur. Phys. J. C}\ }\textbf {\bibinfo {volume} {75}},\
  \bibinfo {pages} {132} (\bibinfo {year} {2015})},\ \Eprint
  {https://arxiv.org/abs/1412.7420} {arXiv:1412.7420 [hep-ph]} \BibitemShut
  {NoStop}%
%%CITATION = ARXIV:1412.7420;%%
\bibitem [{\citenamefont {Suri}\ and\ \citenamefont
  {Yennie}(1972)}]{Suri:1971yx}%
  \BibitemOpen
  \bibfield  {author} {\bibinfo {author} {\bibfnamefont {A.}~\bibnamefont
  {Suri}}\ and\ \bibinfo {author} {\bibfnamefont {D.~R.}\ \bibnamefont
  {Yennie}},\ }\bibfield  {title} {\bibinfo {title} {The space-time
  phenomenology of photon absorption and inelastic electron scattering},\
  }\href {https://doi.org/10.1016/0003-4916(72)90242-4} {\bibfield  {journal}
  {\bibinfo  {journal} {Annals Phys.}\ }\textbf {\bibinfo {volume} {72}},\
  \bibinfo {pages} {243} (\bibinfo {year} {1972})}\BibitemShut {NoStop}%
%%CITATION = APNYA,72,243;%%
\bibitem [{\citenamefont {Fiore}\ \emph {et~al.}(2002)\citenamefont {Fiore},
  \citenamefont {Flachi}, \citenamefont {Jenkovszky}, \citenamefont {Lengyel},\
  and\ \citenamefont {Magas}}]{Fiore:2002re}%
  \BibitemOpen
  \bibfield  {author} {\bibinfo {author} {\bibfnamefont {R.}~\bibnamefont
  {Fiore}}, \bibinfo {author} {\bibfnamefont {A.}~\bibnamefont {Flachi}},
  \bibinfo {author} {\bibfnamefont {L.~L.}\ \bibnamefont {Jenkovszky}},
  \bibinfo {author} {\bibfnamefont {A.~I.}\ \bibnamefont {Lengyel}},\ and\
  \bibinfo {author} {\bibfnamefont {V.~K.}\ \bibnamefont {Magas}},\ }\bibfield
  {title} {\bibinfo {title} {Explicit model realizing parton hadron duality},\
  }\href {https://doi.org/10.1140/epja/i2002-10047-3} {\bibfield  {journal}
  {\bibinfo  {journal} {Eur. Phys. J. A}\ }\textbf {\bibinfo {volume} {15}},\
  \bibinfo {pages} {505} (\bibinfo {year} {2002})},\ \Eprint
  {https://arxiv.org/abs/hep-ph/0206027} {arXiv:hep-ph/0206027 [hep-ph]}
  \BibitemShut {NoStop}%
%%CITATION = HEP-PH/0206027;%%
\bibitem [{\citenamefont {Bosted}\ and\ \citenamefont
  {Christy}(2008)}]{Bosted:2007xd}%
  \BibitemOpen
  \bibfield  {author} {\bibinfo {author} {\bibfnamefont {P.~E.}\ \bibnamefont
  {Bosted}}\ and\ \bibinfo {author} {\bibfnamefont {M.~E.}\ \bibnamefont
  {Christy}},\ }\bibfield  {title} {\bibinfo {title} {Empirical fit to
  inelastic electron-deuteron and electron-neutron resonance region transverse
  cross-sections},\ }\href {https://doi.org/10.1103/PhysRevC.77.065206}
  {\bibfield  {journal} {\bibinfo  {journal} {Phys. Rev. C}\ }\textbf {\bibinfo
  {volume} {77}},\ \bibinfo {pages} {065206} (\bibinfo {year} {2008})},\
  \Eprint {https://arxiv.org/abs/0711.0159} {arXiv:0711.0159 [hep-ph]}
  \BibitemShut {NoStop}%
%%CITATION = ARXIV:0711.0159;%%
\bibitem [{\citenamefont {Szczurek}\ and\ \citenamefont
  {Uleshchenko}(2000)}]{Szczurek:1999rd}%
  \BibitemOpen
  \bibfield  {author} {\bibinfo {author} {\bibfnamefont {A.}~\bibnamefont
  {Szczurek}}\ and\ \bibinfo {author} {\bibfnamefont {V.}~\bibnamefont
  {Uleshchenko}},\ }\bibfield  {title} {\bibinfo {title} {Nonpartonic
  components in the nucleon structure functions at small {$Q^2$} in the broad
  range of {$x$}},\ }\href {https://doi.org/10.1007/s100520000218} {\bibfield
  {journal} {\bibinfo  {journal} {Eur. Phys. J. C}\ }\textbf {\bibinfo {volume}
  {12}},\ \bibinfo {pages} {663} (\bibinfo {year} {2000})},\ \Eprint
  {https://arxiv.org/abs/hep-ph/9904288} {arXiv:hep-ph/9904288 [hep-ph]}
  \BibitemShut {NoStop}%
%%CITATION = HEP-PH/9904288;%%
\bibitem [{\citenamefont {Abramowicz}\ \emph {et~al.}(1991)\citenamefont
  {Abramowicz}, \citenamefont {Levin}, \citenamefont {Levy},\ and\
  \citenamefont {Maor}}]{Abramowicz:1991xz}%
  \BibitemOpen
  \bibfield  {author} {\bibinfo {author} {\bibfnamefont {H.}~\bibnamefont
  {Abramowicz}}, \bibinfo {author} {\bibfnamefont {E.~M.}\ \bibnamefont
  {Levin}}, \bibinfo {author} {\bibfnamefont {A.}~\bibnamefont {Levy}},\ and\
  \bibinfo {author} {\bibfnamefont {U.}~\bibnamefont {Maor}},\ }\bibfield
  {title} {\bibinfo {title} {A parametrization of {$\sigma_T(\gamma^* p)$}
  above the resonance region {$Q^2 \gtrapprox 0$}},\ }\href
  {https://doi.org/10.1016/0370-2693(91)90202-2} {\bibfield  {journal}
  {\bibinfo  {journal} {Phys. Lett. B}\ }\textbf {\bibinfo {volume} {269}},\
  \bibinfo {pages} {465} (\bibinfo {year} {1991})}\BibitemShut {NoStop}%
%%CITATION = PHLTA,B269,465;%%
\bibitem [{\citenamefont {Abramowicz}\ and\ \citenamefont
  {Levy}(1997)}]{Abramowicz:1997ms}%
  \BibitemOpen
  \bibfield  {author} {\bibinfo {author} {\bibfnamefont {H.}~\bibnamefont
  {Abramowicz}}\ and\ \bibinfo {author} {\bibfnamefont {A.}~\bibnamefont
  {Levy}},\ }\bibfield  {title} {\bibinfo {title} {The {ALLM} parameterization
  of $\sigma_{\mathrm{tot}}(\gamma^\ast p)$: An update},\ }\href@noop {} {\
  (\bibinfo {year} {1997})},\ \Eprint {https://arxiv.org/abs/hep-ph/9712415}
  {arXiv:hep-ph/9712415 [hep-ph]} \BibitemShut {NoStop}%
%%CITATION = HEP-PH/9712415;%%
\bibitem [{\citenamefont {Gabbert}\ and\ \citenamefont
  {De~Nardo}(2007)}]{Gabbert:2007bj}%
  \BibitemOpen
  \bibfield  {author} {\bibinfo {author} {\bibfnamefont {D.}~\bibnamefont
  {Gabbert}}\ and\ \bibinfo {author} {\bibfnamefont {L.}~\bibnamefont
  {De~Nardo}},\ }\bibfield  {title} {\bibinfo {title} {New global fit to the
  total photon-proton cross-section {$\sigma_{L+T}$} and to the structure
  function {$F_2$}},\ }in\ \href {https://doi.org/10.3204/proc07-01/44} {\emph
  {\bibinfo {booktitle} {{Proceedings, 15th International Workshop on
  Deep-inelastic scattering and related subjects (DIS 2007). Vol. 1 and 2:
  Munich, Germany}}}}\ (\bibinfo {year} {2007})\ p.\ \bibinfo {pages} {373},\
  \Eprint {https://arxiv.org/abs/0708.3196} {arXiv:0708.3196 [hep-ph]}
  \BibitemShut {NoStop}%
%%CITATION = ARXIV:0708.3196;%%
\bibitem [{\citenamefont {Airapetian}\ \emph {et~al.}(2011)\citenamefont
  {Airapetian} \emph {et~al.}}]{Airapetian:2011nu}%
  \BibitemOpen
  \bibfield  {author} {\bibinfo {author} {\bibfnamefont {A.}~\bibnamefont
  {Airapetian}} \emph {et~al.} (\bibinfo {collaboration} {HERMES}),\ }\bibfield
   {title} {\bibinfo {title} {Inclusive measurements of inelastic electron and
  positron scattering from unpolarized hydrogen and deuterium targets},\ }\href
  {https://doi.org/10.1007/JHEP05(2011)126} {\bibfield  {journal} {\bibinfo
  {journal} {JHEP}\ }\textbf {\bibinfo {volume} {05}},\ \bibinfo {pages}
  {126}},\ \Eprint {https://arxiv.org/abs/1103.5704} {arXiv:1103.5704 [hep-ex]}
  \BibitemShut {NoStop}%
%%CITATION = ARXIV:1103.5704;%%
\bibitem [{\citenamefont {Martin}\ \emph {et~al.}(2009)\citenamefont {Martin},
  \citenamefont {Stirling}, \citenamefont {Thorne},\ and\ \citenamefont
  {Watt}}]{Martin:2009iq}%
  \BibitemOpen
  \bibfield  {author} {\bibinfo {author} {\bibfnamefont {A.~D.}\ \bibnamefont
  {Martin}}, \bibinfo {author} {\bibfnamefont {W.~J.}\ \bibnamefont
  {Stirling}}, \bibinfo {author} {\bibfnamefont {R.~S.}\ \bibnamefont
  {Thorne}},\ and\ \bibinfo {author} {\bibfnamefont {G.}~\bibnamefont {Watt}},\
  }\bibfield  {title} {\bibinfo {title} {Parton distributions for the {LHC}},\
  }\href {https://doi.org/10.1140/epjc/s10052-009-1072-5} {\bibfield  {journal}
  {\bibinfo  {journal} {Eur. Phys. J. C}\ }\textbf {\bibinfo {volume} {63}},\
  \bibinfo {pages} {189} (\bibinfo {year} {2009})},\ \Eprint
  {https://arxiv.org/abs/0901.0002} {arXiv:0901.0002 [hep-ph]} \BibitemShut
  {NoStop}%
%%CITATION = ARXIV:0901.0002;%%
\bibitem [{\citenamefont {Manohar}\ \emph {et~al.}(2017)\citenamefont
  {Manohar}, \citenamefont {Nason}, \citenamefont {Salam},\ and\ \citenamefont
  {Zanderighi}}]{Manohar:2017eqh}%
  \BibitemOpen
  \bibfield  {author} {\bibinfo {author} {\bibfnamefont {A.~V.}\ \bibnamefont
  {Manohar}}, \bibinfo {author} {\bibfnamefont {P.}~\bibnamefont {Nason}},
  \bibinfo {author} {\bibfnamefont {G.~P.}\ \bibnamefont {Salam}},\ and\
  \bibinfo {author} {\bibfnamefont {G.}~\bibnamefont {Zanderighi}},\ }\bibfield
   {title} {\bibinfo {title} {The photon content of the proton},\ }\href
  {https://doi.org/10.1007/JHEP12(2017)046} {\bibfield  {journal} {\bibinfo
  {journal} {JHEP}\ }\textbf {\bibinfo {volume} {12}},\ \bibinfo {pages}
  {046}},\ \Eprint {https://arxiv.org/abs/1708.01256} {arXiv:1708.01256
  [hep-ph]} \BibitemShut {NoStop}%
%%CITATION = ARXIV:1708.01256;%%
\bibitem [{\citenamefont {Benvenuti}\ \emph {et~al.}(1989)\citenamefont
  {Benvenuti} \emph {et~al.}}]{Benvenuti:1989rh}%
  \BibitemOpen
  \bibfield  {author} {\bibinfo {author} {\bibfnamefont {A.~C.}\ \bibnamefont
  {Benvenuti}} \emph {et~al.} (\bibinfo {collaboration} {BCDMS}),\ }\bibfield
  {title} {\bibinfo {title} {A high statistics measurement of the proton
  structure functions {$F_2(x, Q^2)$} and {$R$} from deep inelastic muon
  scattering at high {$Q^2$}},\ }\href
  {https://doi.org/10.1016/0370-2693(89)91637-7} {\bibfield  {journal}
  {\bibinfo  {journal} {Phys. Lett. B}\ }\textbf {\bibinfo {volume} {223}},\
  \bibinfo {pages} {485} (\bibinfo {year} {1989})}\BibitemShut {NoStop}%
%%CITATION = PHLTA,B223,485;%%
\bibitem [{\citenamefont {Gehrmann}\ \emph {et~al.}(1999)\citenamefont
  {Gehrmann}, \citenamefont {Roberts},\ and\ \citenamefont
  {Whalley}}]{Gehrmann:1999xn}%
  \BibitemOpen
  \bibfield  {author} {\bibinfo {author} {\bibfnamefont {T.}~\bibnamefont
  {Gehrmann}}, \bibinfo {author} {\bibfnamefont {R.~G.}\ \bibnamefont
  {Roberts}},\ and\ \bibinfo {author} {\bibfnamefont {M.~R.}\ \bibnamefont
  {Whalley}},\ }\bibfield  {title} {\bibinfo {title} {A compilation of
  structure functions in deep inelastic scattering},\ }\href
  {https://doi.org/10.1088/0954-3899/25/12A/301} {\bibfield  {journal}
  {\bibinfo  {journal} {J. Phys. G}\ }\textbf {\bibinfo {volume} {25}},\
  \bibinfo {pages} {A1} (\bibinfo {year} {1999})}\BibitemShut {NoStop}%
%%CITATION = JPAGA,G25,A1;%%
\bibitem [{\citenamefont {Abe}\ \emph {et~al.}(1999)\citenamefont {Abe} \emph
  {et~al.}}]{Abe:1998ym}%
  \BibitemOpen
  \bibfield  {author} {\bibinfo {author} {\bibfnamefont {K.}~\bibnamefont
  {Abe}} \emph {et~al.} (\bibinfo {collaboration} {E143}),\ }\bibfield  {title}
  {\bibinfo {title} {Measurements of {$R = \sigma_L / \sigma_T$} for {$0.03 <
  \xbj < 0.1$} and fit to world data},\ }\href
  {https://doi.org/10.1016/S0370-2693(99)00244-0} {\bibfield  {journal}
  {\bibinfo  {journal} {Phys. Lett. B}\ }\textbf {\bibinfo {volume} {452}},\
  \bibinfo {pages} {194} (\bibinfo {year} {1999})},\ \Eprint
  {https://arxiv.org/abs/hep-ex/9808028} {arXiv:hep-ex/9808028 [hep-ex]}
  \BibitemShut {NoStop}%
%%CITATION = HEP-EX/9808028;%%
\bibitem [{\citenamefont {Whitlow}\ \emph {et~al.}(1990)\citenamefont
  {Whitlow}, \citenamefont {Rock}, \citenamefont {Bodek}, \citenamefont
  {Riordan},\ and\ \citenamefont {Dasu}}]{Whitlow:1990gk}%
  \BibitemOpen
  \bibfield  {author} {\bibinfo {author} {\bibfnamefont {L.~W.}\ \bibnamefont
  {Whitlow}}, \bibinfo {author} {\bibfnamefont {S.}~\bibnamefont {Rock}},
  \bibinfo {author} {\bibfnamefont {A.}~\bibnamefont {Bodek}}, \bibinfo
  {author} {\bibfnamefont {E.~M.}\ \bibnamefont {Riordan}},\ and\ \bibinfo
  {author} {\bibfnamefont {S.}~\bibnamefont {Dasu}},\ }\bibfield  {title}
  {\bibinfo {title} {A precise extraction of {$R = \sigma_L / \sigma_T$} from a
  global analysis of the {SLAC} deep inelastic $ep$ and $ed$ scattering
  cross-sections},\ }\href {https://doi.org/10.1016/0370-2693(90)91176-C}
  {\bibfield  {journal} {\bibinfo  {journal} {Phys. Lett. B}\ }\textbf
  {\bibinfo {volume} {250}},\ \bibinfo {pages} {193} (\bibinfo {year}
  {1990})}\BibitemShut {NoStop}%
%%CITATION = PHLTA,B250,193;%%
\bibitem [{\citenamefont {Sibirtsev}\ and\ \citenamefont
  {Blunden}(2013)}]{Sibirtsev:2013cga}%
  \BibitemOpen
  \bibfield  {author} {\bibinfo {author} {\bibfnamefont {A.}~\bibnamefont
  {Sibirtsev}}\ and\ \bibinfo {author} {\bibfnamefont {P.~G.}\ \bibnamefont
  {Blunden}},\ }\bibfield  {title} {\bibinfo {title} {{$Q^2$} evolution of the
  electric and magnetic polarizabilities of the proton},\ }\href
  {https://doi.org/10.1103/PhysRevC.88.065202} {\bibfield  {journal} {\bibinfo
  {journal} {Phys. Rev. C}\ }\textbf {\bibinfo {volume} {88}},\ \bibinfo
  {pages} {065202} (\bibinfo {year} {2013})},\ \Eprint
  {https://arxiv.org/abs/1311.6482} {arXiv:1311.6482 [nucl-th]} \BibitemShut
  {NoStop}%
%%CITATION = ARXIV:1311.6482;%%
\bibitem [{\citenamefont {Galassi}\ \emph {et~al.}(2009)\citenamefont {Galassi}
  \emph {et~al.}}]{Gough:2009:GSL:1538674}%
  \BibitemOpen
  \bibfield  {author} {\bibinfo {author} {\bibfnamefont {M.}~\bibnamefont
  {Galassi}} \emph {et~al.},\ }\href {http://www.gnu.org/software/gsl/} {\emph
  {\bibinfo {title} {{GNU Scientific Library} Reference Manual}}},\ \bibinfo
  {edition} {3rd}\ ed.\ (\bibinfo  {publisher} {Network Theory Ltd.},\ \bibinfo
  {year} {2009})\BibitemShut {NoStop}%
\bibitem [{\citenamefont {Lepage}(1978)}]{Lepage:1977sw}%
  \BibitemOpen
  \bibfield  {author} {\bibinfo {author} {\bibfnamefont {G.~P.}\ \bibnamefont
  {Lepage}},\ }\bibfield  {title} {\bibinfo {title} {A new algorithm for
  adaptive multidimensional integration},\ }\href
  {https://doi.org/10.1016/0021-9991(78)90004-9} {\bibfield  {journal}
  {\bibinfo  {journal} {J. Comput. Phys.}\ }\textbf {\bibinfo {volume} {27}},\
  \bibinfo {pages} {192} (\bibinfo {year} {1978})}\BibitemShut {NoStop}%
%%CITATION = JCTPA,27,192;%%
\bibitem [{\citenamefont {Press}\ and\ \citenamefont
  {Farrar}(1990)}]{Press:1989vk}%
  \BibitemOpen
  \bibfield  {author} {\bibinfo {author} {\bibfnamefont {W.~H.}\ \bibnamefont
  {Press}}\ and\ \bibinfo {author} {\bibfnamefont {G.~R.}\ \bibnamefont
  {Farrar}},\ }\bibfield  {title} {\bibinfo {title} {Recursive stratified
  sampling for multidimensional {Monte Carlo} integration},\ }\href
  {https://doi.org/10.1063/1.4822899} {\bibfield  {journal} {\bibinfo
  {journal} {Computers in Physics}\ }\textbf {\bibinfo {volume} {4}},\ \bibinfo
  {pages} {190} (\bibinfo {year} {1990})}\BibitemShut {NoStop}%
%%CITATION = CFA-3010;%%
\bibitem [{\citenamefont {Dobbs}\ and\ \citenamefont
  {Hansen}(2001)}]{Dobbs:2001ck}%
  \BibitemOpen
  \bibfield  {author} {\bibinfo {author} {\bibfnamefont {M.}~\bibnamefont
  {Dobbs}}\ and\ \bibinfo {author} {\bibfnamefont {J.~B.}\ \bibnamefont
  {Hansen}},\ }\bibfield  {title} {\bibinfo {title} {The {HepMC} {C++} {Monte
  Carlo} event record for high energy physics},\ }\href
  {https://doi.org/10.1016/S0010-4655(00)00189-2} {\bibfield  {journal}
  {\bibinfo  {journal} {Comput. Phys. Commun.}\ }\textbf {\bibinfo {volume}
  {134}},\ \bibinfo {pages} {41} (\bibinfo {year} {2001})}\BibitemShut
  {NoStop}%
%%CITATION = CPHCB,134,41;%%
\bibitem [{\citenamefont {Alwall}\ \emph {et~al.}(2007)\citenamefont {Alwall}
  \emph {et~al.}}]{Alwall:2006yp}%
  \BibitemOpen
  \bibfield  {author} {\bibinfo {author} {\bibfnamefont {J.}~\bibnamefont
  {Alwall}} \emph {et~al.},\ }\bibfield  {title} {\bibinfo {title} {A standard
  format for {Les Houches} event files},\ }\bibfield  {booktitle} {\emph
  {\bibinfo {booktitle} {{Monte Carlos} for the {LHC}: A Workshop on the Tools
  for {LHC} Event Simulation ({MC4LHC}) Geneva, Switzerland, July 17-16,
  2006}},\ }\href {https://doi.org/10.1016/j.cpc.2006.11.010} {\bibfield
  {journal} {\bibinfo  {journal} {Comput. Phys. Commun.}\ }\textbf {\bibinfo
  {volume} {176}},\ \bibinfo {pages} {300} (\bibinfo {year} {2007})},\ \Eprint
  {https://arxiv.org/abs/hep-ph/0609017} {arXiv:hep-ph/0609017 [hep-ph]}
  \BibitemShut {NoStop}%
%%CITATION = HEP-PH/0609017;%%
\bibitem [{\citenamefont {Brun}\ and\ \citenamefont
  {Rademakers}(1997)}]{Brun:1997pa}%
  \BibitemOpen
  \bibfield  {author} {\bibinfo {author} {\bibfnamefont {R.}~\bibnamefont
  {Brun}}\ and\ \bibinfo {author} {\bibfnamefont {F.}~\bibnamefont
  {Rademakers}},\ }\bibfield  {title} {\bibinfo {title} {{ROOT}: An object
  oriented data analysis framework},\ }\bibfield  {booktitle} {\emph {\bibinfo
  {booktitle} {New computing techniques in physics research V. Proceedings, 5th
  International Workshop, {AIHENP} '96, Lausanne, Switzerland, September 2-6,
  1996}},\ }\href {https://doi.org/10.1016/S0168-9002(97)00048-X} {\bibfield
  {journal} {\bibinfo  {journal} {Nucl. Instrum. Meth. A}\ }\textbf {\bibinfo
  {volume} {389}},\ \bibinfo {pages} {81} (\bibinfo {year} {1997})}\BibitemShut
  {NoStop}%
%%CITATION = NUIMA,A389,81;%%
\bibitem [{\citenamefont {Sj{\"{o}}strand}\ \emph {et~al.}(2015)\citenamefont
  {Sj{\"{o}}strand}, \citenamefont {Ask}, \citenamefont {Christiansen},
  \citenamefont {Corke}, \citenamefont {Desai}, \citenamefont {Ilten},
  \citenamefont {Mrenna}, \citenamefont {Prestel}, \citenamefont {Rasmussen},\
  and\ \citenamefont {Skands}}]{Sjostrand:2014zea}%
  \BibitemOpen
  \bibfield  {author} {\bibinfo {author} {\bibfnamefont {T.}~\bibnamefont
  {Sj{\"{o}}strand}}, \bibinfo {author} {\bibfnamefont {S.}~\bibnamefont
  {Ask}}, \bibinfo {author} {\bibfnamefont {J.~R.}\ \bibnamefont
  {Christiansen}}, \bibinfo {author} {\bibfnamefont {R.}~\bibnamefont {Corke}},
  \bibinfo {author} {\bibfnamefont {N.}~\bibnamefont {Desai}}, \bibinfo
  {author} {\bibfnamefont {P.}~\bibnamefont {Ilten}}, \bibinfo {author}
  {\bibfnamefont {S.}~\bibnamefont {Mrenna}}, \bibinfo {author} {\bibfnamefont
  {S.}~\bibnamefont {Prestel}}, \bibinfo {author} {\bibfnamefont {C.~O.}\
  \bibnamefont {Rasmussen}},\ and\ \bibinfo {author} {\bibfnamefont {P.~Z.}\
  \bibnamefont {Skands}},\ }\bibfield  {title} {\bibinfo {title} {An
  introduction to {PYTHIA} 8.2},\ }\href
  {https://doi.org/10.1016/j.cpc.2015.01.024} {\bibfield  {journal} {\bibinfo
  {journal} {Comput. Phys. Commun.}\ }\textbf {\bibinfo {volume} {191}},\
  \bibinfo {pages} {159} (\bibinfo {year} {2015})},\ \Eprint
  {https://arxiv.org/abs/1410.3012} {arXiv:1410.3012 [hep-ph]} \BibitemShut
  {NoStop}%
%%CITATION = ARXIV:1410.3012;%%
\bibitem [{\citenamefont {Bahr}\ \emph {et~al.}(2008)\citenamefont {Bahr} \emph
  {et~al.}}]{Bahr:2008pv}%
  \BibitemOpen
  \bibfield  {author} {\bibinfo {author} {\bibfnamefont {M.}~\bibnamefont
  {Bahr}} \emph {et~al.},\ }\bibfield  {title} {\bibinfo {title} {{Herwig++}
  physics and manual},\ }\href {https://doi.org/10.1140/epjc/s10052-008-0798-9}
  {\bibfield  {journal} {\bibinfo  {journal} {Eur. Phys. J. C}\ }\textbf
  {\bibinfo {volume} {58}},\ \bibinfo {pages} {639} (\bibinfo {year} {2008})},\
  \Eprint {https://arxiv.org/abs/0803.0883} {arXiv:0803.0883 [hep-ph]}
  \BibitemShut {NoStop}%
%%CITATION = ARXIV:0803.0883;%%
\bibitem [{\citenamefont {Bellm}\ \emph {et~al.}(2016)\citenamefont {Bellm}
  \emph {et~al.}}]{Bellm:2015jjp}%
  \BibitemOpen
  \bibfield  {author} {\bibinfo {author} {\bibfnamefont {J.}~\bibnamefont
  {Bellm}} \emph {et~al.},\ }\bibfield  {title} {\bibinfo {title} {{Herwig
  7.0/Herwig++ 3.0} release note},\ }\href
  {https://doi.org/10.1140/epjc/s10052-016-4018-8} {\bibfield  {journal}
  {\bibinfo  {journal} {Eur. Phys. J. C}\ }\textbf {\bibinfo {volume} {76}},\
  \bibinfo {pages} {196} (\bibinfo {year} {2016})},\ \Eprint
  {https://arxiv.org/abs/1512.01178} {arXiv:1512.01178 [hep-ph]} \BibitemShut
  {NoStop}%
%%CITATION = ARXIV:1512.01178;%%
\bibitem [{\citenamefont {Sirunyan}\ \emph {et~al.}(2018)\citenamefont
  {Sirunyan} \emph {et~al.}}]{Cms:2018het}%
  \BibitemOpen
  \bibfield  {author} {\bibinfo {author} {\bibfnamefont {A.~M.}\ \bibnamefont
  {Sirunyan}} \emph {et~al.} (\bibinfo {collaboration} {CMS, TOTEM}),\
  }\bibfield  {title} {\bibinfo {title} {Observation of proton-tagged, central
  (semi)exclusive production of high-mass lepton pairs in {pp} collisions at 13
  {TeV} with the {CMS-TOTEM} precision proton spectrometer},\ }\href
  {https://doi.org/10.1007/JHEP07(2018)153} {\bibfield  {journal} {\bibinfo
  {journal} {JHEP}\ }\textbf {\bibinfo {volume} {07}},\ \bibinfo {pages}
  {153}},\ \Eprint {https://arxiv.org/abs/1803.04496} {arXiv:1803.04496
  [hep-ex]} \BibitemShut {NoStop}%
%%CITATION = ARXIV:1803.04496;%%
\bibitem [{\citenamefont {Forthomme}\ \emph {et~al.}(2019)\citenamefont
  {Forthomme}, \citenamefont {{\L{}}uszczak}, \citenamefont {Sch{\"{a}}fer},\
  and\ \citenamefont {Szczurek}}]{Forthomme:2018sxa}%
  \BibitemOpen
  \bibfield  {author} {\bibinfo {author} {\bibfnamefont {L.}~\bibnamefont
  {Forthomme}}, \bibinfo {author} {\bibfnamefont {M.}~\bibnamefont
  {{\L{}}uszczak}}, \bibinfo {author} {\bibfnamefont {W.}~\bibnamefont
  {Sch{\"{a}}fer}},\ and\ \bibinfo {author} {\bibfnamefont {A.}~\bibnamefont
  {Szczurek}},\ }\bibfield  {title} {\bibinfo {title} {Rapidity gap survival
  factors caused by remnant fragmentation for {$W^+ W^-$} pair production via
  {$\gamma^*\gamma^* \to W^+ W^-$} subprocess with photon transverse momenta},\
  }\href {https://doi.org/10.1016/j.physletb.2018.12.018} {\bibfield  {journal}
  {\bibinfo  {journal} {Phys. Lett. B}\ }\textbf {\bibinfo {volume} {789}},\
  \bibinfo {pages} {300} (\bibinfo {year} {2019})},\ \Eprint
  {https://arxiv.org/abs/1805.07124} {arXiv:1805.07124 [hep-ph]} \BibitemShut
  {NoStop}%
%%CITATION = ARXIV:1805.07124;%%
\bibitem [{\citenamefont {Harland-Lang}\ \emph {et~al.}(2016)\citenamefont
  {Harland-Lang}, \citenamefont {Khoze},\ and\ \citenamefont
  {Ryskin}}]{Harland-Lang:2016apc}%
  \BibitemOpen
  \bibfield  {author} {\bibinfo {author} {\bibfnamefont {L.~A.}\ \bibnamefont
  {Harland-Lang}}, \bibinfo {author} {\bibfnamefont {V.~A.}\ \bibnamefont
  {Khoze}},\ and\ \bibinfo {author} {\bibfnamefont {M.~G.}\ \bibnamefont
  {Ryskin}},\ }\bibfield  {title} {\bibinfo {title} {The photon {PDF} in events
  with rapidity gaps},\ }\href {https://doi.org/10.1140/epjc/s10052-016-4100-2}
  {\bibfield  {journal} {\bibinfo  {journal} {Eur. Phys. J. C}\ }\textbf
  {\bibinfo {volume} {76}},\ \bibinfo {pages} {255} (\bibinfo {year} {2016})},\
  \Eprint {https://arxiv.org/abs/1601.03772} {arXiv:1601.03772 [hep-ph]}
  \BibitemShut {NoStop}%
%%CITATION = ARXIV:1601.03772;%%
\end{thebibliography}%

\end{document}